\documentclass[floatfix, reprint,
amsmath,amssymb,
 aps, superscriptaddress,longbibliography]{revtex4-2}
\usepackage[T1]{fontenc}
\usepackage{braket}
\usepackage{graphicx}
\usepackage{bbold}
\usepackage{siunitx}
\usepackage[noend]{algpseudocode}
\usepackage{dcolumn}
\usepackage{bm}
\usepackage{hyperref}
\usepackage{placeins}
\begin{document}


\title{g-tensor Optimization in Ge/SiGe Quantum Dots}

\author{Aram Shojaei}
\affiliation{QuTech and Kavli Institute of Nanoscience, Delft University of Technology, 2628 CJ Delft, the Netherlands}
\author{Edmondo Valvo}
\affiliation{QuTech and Kavli Institute of Nanoscience, Delft University of Technology, 2628 CJ Delft, the Netherlands}
\author{Maximilian Rimbach-Russ}
\affiliation{QuTech and Kavli Institute of Nanoscience, Delft University of Technology, 2628 CJ Delft, the Netherlands}
\author{Eliska Greplova}
\affiliation{QuTech and Kavli Institute of Nanoscience, Delft University of Technology, 2628 CJ Delft, the Netherlands}
\author{Ana Silva}
\affiliation{QuTech and Kavli Institute of Nanoscience, Delft University of Technology, 2628 CJ Delft, the Netherlands}

 \date{\today}

\begin{abstract}
Planar germanium heterostructures hosting hole-spin qubits are among the leading platforms for scalable semiconductor-based quantum computing. Yet, device performance is hindered by significant quantum dot variability, which leads to uncertainty in qubit energy levels and random orientations of the spin quantization axis. Tailored control of the $g$-tensor offers a strategy to overcome these limitations and achieve more reliable qubit operations. Here, we introduce a flexible optimization framework for engineering $g$-tensor properties. As a benchmark, we numerically obtain the optimal reshaping of the out-of-plane potential in a SiGe–Ge–SiGe quantum well to suppress the in-plane $g$-tensor components and realize the recently proposed gapless single-spin qubit encoding. This reshaping is achieved through heterostructure engineering, specifically by adjusting the silicon concentration within the quantum well, though the framework remains readily adaptable to alternative design objectives. Our results provide practical design principles for improving the tunability of the spin response, paving the way towards large-scale germanium-based quantum computers.

\end{abstract}

\maketitle

\section{Introduction}

Solid-state spin qubits in semiconductor quantum dots are a promising platform for achieving scalable quantum processors~\cite{Zwerver_2023,Maurand2016CMOSSpinQubit,Gonzalez_Zalba_2021,Takeda_2022,Xue2022SurfaceCode,Philips_2022,Loss_1998,Burkard2023_RMP_SpinQubits,Kloeffel2013_ARCMP,Vandersypen_2017,HamBen2025,ZwerKraheWat2022}. In particular, holes confined in germanium (Ge) quantum dots offer several favorable features for qubit implementation~\cite{vorreiter2025precisionhighspeedquantumlogic,John2025RobustGe10Qubit,Stehouwer_2025,Zhang_2025,Lawrie_2023,Liu_2023,WanGanGao2022,Hendrickx_2021,Scappucci_2021,Watzinger2018NatComm, Hendrickx_2020,Jirovec_2021}. These features include a strong spin–orbit coupling (SOC) that enables fast all-electric control~\cite{Froning2021_NatNanotech}; a weak hyperfine interaction that enhances qubit coherence times~\cite{Fang2023HoleSpinReview,WanGanGao2022,Scappucci_2021}; and the lack of valley degeneracy, preventing additional leakage channels through nearby degenerate energy levels~\cite{Terrazos_2021}. Furthermore, the small effective mass of holes in Ge enables larger quantum dots, thereby reducing fabrication constraints and ultimately aiding device scalability~\cite{Terrazos_2021,MauRodMena2025}.

The intrinsically strong SOC of hole states in group-IV semiconductors, such as Ge and Si, is associated with a highly anisotropic $g$-tensor \cite{Hendrickx_2024,Qvist_2022,jin2024probinggtensorreproducibilityspinorbit}. While this anisotropy can aid in qubit control~\cite{Venitucci_2018,Hendrickx_2024}, it also further enhances the sensitivity to both electrostatic variability and to small changes in quantum-dot confinement~\cite{Hendrickx_2024}. Hence, exploiting design principles that explicitly engineer and controllably tune the $g$-tensor's anisotropy is highly desirable for attaining operational sweet spots~\cite{Hendrickx_2024}. Beyond $g$-tensor engineering, both electron and hole spin qubits offer flexibility for baseband control, enabling hopping-based control~\cite{Wang_2024,Unseld2025Baseband2D}, gapless single-spin qubit encoding~\cite{Rimbach-Russ2025}, and multi-spin qubit encodings such as singlet-triplet~\cite{Petta2005_Science,Jirovec_2021,Zhang_2025} and exchange-only~\cite{DiVincenzo_2000,Russ_2017, Burkard2023_RMP_SpinQubits,Weinstein_2023,Bosco_2026}.

A central challenge for hole-spin qubits in planar germanium is their pronounced variability \cite{Valvo2025ElectricallyTuneable,Scappucci_2021,Hendrickx_2024}: owing to strong spin--orbit coupling, the $g$-tensor is highly sensitive to the confinement potential, strain, electrostatics, and disorder. Consequently, large device-to-device (or even cooldown-to-cooldown) fluctuations can arise even in nominally identical heterostructures, limiting reproducibility. Overcoming this variability requires not only sufficient tunability of the relevant spin parameters, but also a systematic strategy to steer individual devices towards desired operating points. This naturally motivates optimization-based and auto-tuning approaches, which have proven highly successful for calibrating charge configurations and quantum-dot parameters in semiconductor devices~\cite{katiraee2025unified,Baart2016,van_Diepen_2018,Zwolak2020,vandriel2024crossplatformautonomouscontrolminimal, KochDrielBordin2023, DurKoskiLandGrep2020}. These approaches rely on closed-loop algorithms that systematically adjust gate voltages to reach desired operating points, such as single-electron occupation and tunnel couplings. Extending this paradigm to spin degrees of freedom is, however, nontrivial, as spin properties are generally only indirectly tunable and exhibit strong sensitivity to quasi-static variations in device confinement, strain, and electrostatics \cite{Hendrickx_2020, Kloeffel2013_ARCMP, Scappucci_2021}. Such an extension becomes feasible only if the system provides sufficiently strong, smooth, and reproducible control degrees of freedom that directly influence the relevant spin parameters.

In this work, we show that targeted Si incorporation modifies the vertical confinement potential, enhancing the tunability of the hole-spin $g$-tensor and enabling optimization and auto-tuning-based control of spin qubits. By discretizing the growth direction and defining a suitable cost function for the desired Zeeman response, we formulate a general framework that allows the $g$-tensor to be systematically optimized using automated search algorithms. As a benchmark, we apply this method to the realization of the recently proposed gapless single-spin encoding~\cite{Rimbach-Russ2025}, which requires suppressing the in-plane $g$-tensor components of a single Kramers doublet. Vertical potential engineering provides robust access to this regime. Using the covariance matrix adaptation evolution strategy (CMA-ES)~\cite{hansen2006cma, nomura2024cmaessimplepractical}, we identify Si profiles that minimize the in-plane $g$-tensor components. Once the Si profile is fixed and optimized, residual fine-tuning (e.g., to further bring $g_{xx}$ closer to the target $g_{xx}=0$) can be performed by adjusting the in-plane confinement lengths $L_x$ and $L_y$. While optimization towards the gapless single-spin encoding serves here as a particular test case, the framework itself is general and can be readily adapted to target other $g$-tensors or qubit properties. Additionally, we show that this optimal solution remains stable under variations in the out-of-plane electric field, thereby helping mitigate device-to-device variability. Our approach complements strain-engineering methods that modulate the Zeeman response through uniaxial or shear strain~\cite{mauro2025strainengineeringgegesispin,AbadilloUriel2023_PRL}, and can be combined with post-fabrication electrostatic squeezing to further tune the $g$-tensor~\cite{Valvo2025ElectricallyTuneable}.

This paper is organized as follows. In Sec.~\ref{sec:methods}, we introduce our general optimization framework, which combines a low-energy effective spin-qubit model with the covariance-matrix adaptation evolution strategy (CMA-ES), a gradient-free optimization algorithm. We then illustrate the approach using a concrete example, where we target the suppression of the in-plane $g$-tensor components. Sec.~\ref{sec:results} presents the optimized Si profiles for the selected example and evaluates their robustness under quasi-static variations of the vertical electric field. In Sec.~\ref{sec:knobs}, we outline the simulation parameters that can be readily adjusted to accommodate alternative optimization objectives. Finally, Sec.~\ref{sec:discussion} discusses the results and outlines potential extensions.

\section{Methods}
\label{sec:methods}
In this section, we introduce our general optimization framework and demonstrate its applicability to engineer a targeted $g$-tensor response in a strained Ge/SiGe hole quantum well. Sec.~\ref{sec:spin_qubit_model} introduces the spin-qubit model and defines the effective $g$-tensor, while establishing the optimization problem. Sec.~\ref{sec:benchmark_gapless} then illustrates the approach applied to a strained Ge/SiGe hole quantum well and identifies the physical parameters controlling the in-plane $g$-tensor response. Finally, Sec.~\ref{sec:cmaes_opt} describes the CMA-ES procedure used to optimize the target Zeeman response in our benchmark example.

\subsection{Spin qubit model}
\label{sec:spin_qubit_model}

We consider a generic semiconductor heterostructure described by a multiband $\mathbf{k}\!\cdot\!\mathbf{p}$ Hamiltonian \cite{Winkler2003_Book}. At zero magnetic field ($\mathbf{B}=\mathbf{0}$), the total Hamiltonian is
\begin{equation}
    H_0(\mathbf{k})=H_{\mathbf{k} \cdot \mathbf{p}} + H_{\text{strain}} + V(\mathbf{r};\boldsymbol{s},E_z),
\label{baseline_H}
\end{equation}
where $H_{\mathbf{k} \cdot \mathbf{p}}$ describes the band mixing near $\mathbf{k}=\mathbf 0$ (for example, a Luttinger-Kohn Hamiltonian describing the $J=3/2$ valence-band manifold \cite{Winkler2003_Book}) and $H_{\text{strain}}$ captures strain effects (e.g., via the Bir-Pikus Hamiltonian \cite{Winkler2003_Book}). The term $V(\mathbf{r};\boldsymbol{s},E_z)$ denotes the quantum dot's electrostatic confinement potential and is parameterized by a set of design variables $\mathbf{s}$, which can include, for example, local material composition or other structural parameters. Here, $\mathbf{r}$ denotes the spatial coordinate of the carrier.

For planar quantum dots with $L_z \ll L_x,L_y$, the confinement potential is taken to be separable \cite{Martinez2022_NonSeparableFields, Wang2024_npjQI}:
\begin{equation}
V(\mathbf{r;\boldsymbol{s}},E_z)=V_{\perp}(z;\boldsymbol{s},E_z) + V_{\parallel}(x,y),
\label{total_potential}
\end{equation}
where $V_{\perp}(z;\boldsymbol{s},E_z)$ describes the full growth-direction potential, $V_{\parallel}(x,y)$ denotes the lateral confinement of the quantum dot (e.g., 2D harmonic potential), and $E_{\text{z}}$ is the average vertical electric field.

Assuming time-reversal symmetry at $\mathbf{B}=\mathbf{0}$, we encode the qubit in the lowest-energy Kramers doublet of $H_0$. The states $\{ \ket{\mathbb{0}},\ket{\mathbb{1}}\}$ define an orthonormal basis spanning this ground state doublet, and we introduce the projector $P_0=\ket{\mathbb{0}}\bra{\mathbb{0}}+\ket{\mathbb{1}}\bra{\mathbb{1}}$ \cite{Venitucci_2018}. With the qubit basis fixed at $\mathbf{B}=\mathbf 0$, the effects of a weak magnetic field $\mathbf{B}$ are captured by two contributions: the Zeeman interaction, $H_Z(\mathbf{B})$, and the orbital coupling, which enters through the minimal substitution $\mathbf{k}\rightarrow \mathbf{k} + \tfrac{e}{\hbar}\mathbf{A}$ in $H_{\mathbf{k} \cdot \mathbf{p}}$, where $\mathbf{A}$ is the magnetic vector potential satisfying $\nabla\times\mathbf{A}=\mathbf{B}$. The full Hamiltonian is then given by
\begin{equation}
H(\mathbf{B}) \;=\; H_0 \;+\; H_1(\mathbf{B}) \;+\; \mathcal{O}(B^2),
\label{eq:H_B_expand}
\end{equation}
\begin{equation}
    H_1(\mathbf{B}) \equiv H_Z(\mathbf{B}) + H_{\rm orb}(\mathbf{B}),
\end{equation}
where $H_{\rm orb}(\mathbf{B})$ collects the linear terms in $\mathbf{B}$ generated by the substitution in $H_{\mathbf{k}\cdot\mathbf{p}}$. Here $H_1(\mathbf{B})$ is treated perturbatively to first order in $\mathbf{B}$.

Projecting  $H_1(\mathbf{B})$ onto the ground state doublet subspace yields a $2\times2$ effective Hamiltonian \cite{Venitucci_2018, Crippa_2018},
\begin{equation}
H_{\rm eff}^{(2\times2)}(\mathbf B)
=
P_0\,H_1(\mathbf{B})\,P_0 = \frac{\mu_B}{2}
\bm{\sigma} \cdot\,\mathbf{g}\,\cdot\mathbf B,
\label{eq:Heff_proj}
\end{equation}
that defines the $g$-tensor of the lowest Kramers doublet in the chosen doublet basis $\{\ket{\mathbb{0}},\ket{\mathbb{1}}\}$.

Instead of explicitly computing the Kramers doublet, which may be computationally demanding in multiband models, one may equivalently construct an effective low-energy subspace using a Schrieffer–Wolff (SW) transformation \cite{Lowdin1951}. Starting from a convenient two-state subspace $\mathcal{P}=\mathrm{span}\{\ket{0},\ket{1}\}$ and its complement $\mathcal{Q}=\mathcal{P}^\perp$, a unitary transformation is chosen such that
\begin{equation}
\tilde H_0 = U_{\rm SW}^\dagger H_0\,U_{\rm SW}
\end{equation}
is (approximately) block diagonal in the $\mathcal{P}\oplus \mathcal{Q}$ decomposition at $\mathbf B=\mathbf 0$. The SW dressed states $|\tilde{0} \rangle=U^{\dagger}_{SW} \ket{0}$ and $|\tilde{1}\rangle=  U^{\dagger}_{SW} \ket{1}$ then approximate the low-energy eigenspace of $H_0$. Using the same unitary for the linear-in-$\mathbf B$ term,
\begin{equation}
\tilde H_1(\mathbf B)=U_{\rm SW}^\dagger H_1(\mathbf B)\,U_{\rm SW},
\end{equation}
the effective qubit Hamiltonian is obtained by projecting onto the SW-dressed two-state subspace,
\begin{equation}
H_{\rm eff}^{(2\times2)}(\mathbf B)=P_{\rm SW}\,\tilde H_1(\mathbf B)\,P_{\rm SW}, \qquad P_{\rm SW}=\ket{\tilde{0}}\bra{\tilde{0}}+\ket{\tilde{1}}\bra{\tilde{1}}
\label{eq:Heff_g_tensor_def}
\end{equation}
and $\mathbf g$ is extracted by matching again to the right-hand side of Eq.~\eqref{eq:Heff_proj}~\cite{Sarkar2023_PRB}. While the explicit representation of the $g$-tensor depends on the chosen basis and therefore can differ between the Kramers doublet ${\ket{\mathbb{0}}, \ket{\mathbb{1}}}$ and the Schrieffer–Wolff–dressed basis ${\ket{\tilde{0}}, \ket{\tilde{1}}}$, physical observables, such as the Zeeman splitting projected along the magnetic-field direction, remain invariant~\cite{Venitucci_2018}.

The $g$-tensor depends on the design parameters $\mathbf{s}$ through the confinement potential $V_\perp(z;\mathbf s,E_z)$ appearing in $H_0$. This dependence naturally defines an optimization problem over $\mathbf{s}$, aimed at identifying parameter configurations for which the resulting $g$-tensor most closely matches a prescribed target matrix. For our benchmark, $\boldsymbol{s}$ corresponds to the vertical material profile, specifically the silicon concentration across the quantum well, which is fixed at fabrication and provides a robust baseline Zeeman response.

To achieve a desired Zeeman response, we recast the problem into minimizing a scalar loss function,
\begin{equation}
\mathcal{L}(\boldsymbol{s}) \;=\; \mathcal{G}\;\!\big(\mathbf{g}(\boldsymbol{s})\big) \;+\; \mathcal{P}(\boldsymbol{s}) \; ,
\label{eq:loss_general}
\end{equation}
where $\mathcal{G}$ describes the $g$-tensor objective, such as suppression of specific components or enforcement of a desired anisotropy, and $\mathcal{P}$ penalizes physically infeasible or undesired solutions. In this framework, the target Zeeman response is achieved by finding the set of design parameters $\boldsymbol{s}$ that minimizes $\mathcal{L}$. The form of the penalty term $\mathcal{P}$ is flexible and can incorporate any physical or fabrication constraints relevant to a given device. For example, it can prevent loss of confinement or wave function leakage outside the intended region. In buried quantum-well spin qubits, such as germanium hole or SiGe electron qubits, one may, for example, impose an energy-based confinement criterion by taking the oxide-side band edge as a reference level and penalizing candidate profiles for which the relevant low-energy state approaches or exceeds that reference level. This enforces that the state remains energetically bound below the oxide interface and reduces the risk of wave-function transfer towards the oxide interface.

The optimization task therefore consists in identifying the parameter set
\begin{equation}
\boldsymbol{s}^*=\arg\min_{\boldsymbol{s}}\mathcal{L}(\boldsymbol{s})\; .    
\end{equation}
Depending on the specific choice of design variables $\boldsymbol{s}$, this minimization may involve a high-dimensional and nontrivial parameter landscape. In such cases, employing automated search algorithms enables efficient navigation of complex landscapes and the identification of improved candidate solutions. In the present work, we use the covariance-matrix adaptation evolution strategy (CMA-ES), a derivative-free optimization method that is well suited to this setting because it can handle nonlinear, high-dimensional landscapes without requiring gradient information, while iteratively adapting its sampling distribution towards lower-loss regions~\cite{hansen2006cma,nomura2024cmaessimplepractical,katiraee2025unified}.

Once the optimized design variables $\boldsymbol{s}$ are fixed, the resulting target $g$-tensor can be further adjusted through experimentally accessible control parameters. In this way, the optimization over $\boldsymbol{s}$ defines a fabrication-level baseline, while post-fabrication tuning enables refinement of the Zeeman response around the desired operating point. In the present work, we later illustrate this strategy using the lateral confinement lengths $L_x$ and $L_y$.

\subsection{Benchmark application: gapless spin qubit}
\label{sec:benchmark_gapless}

\begin{figure}[t]
  \centering
  \includegraphics[width=\columnwidth]{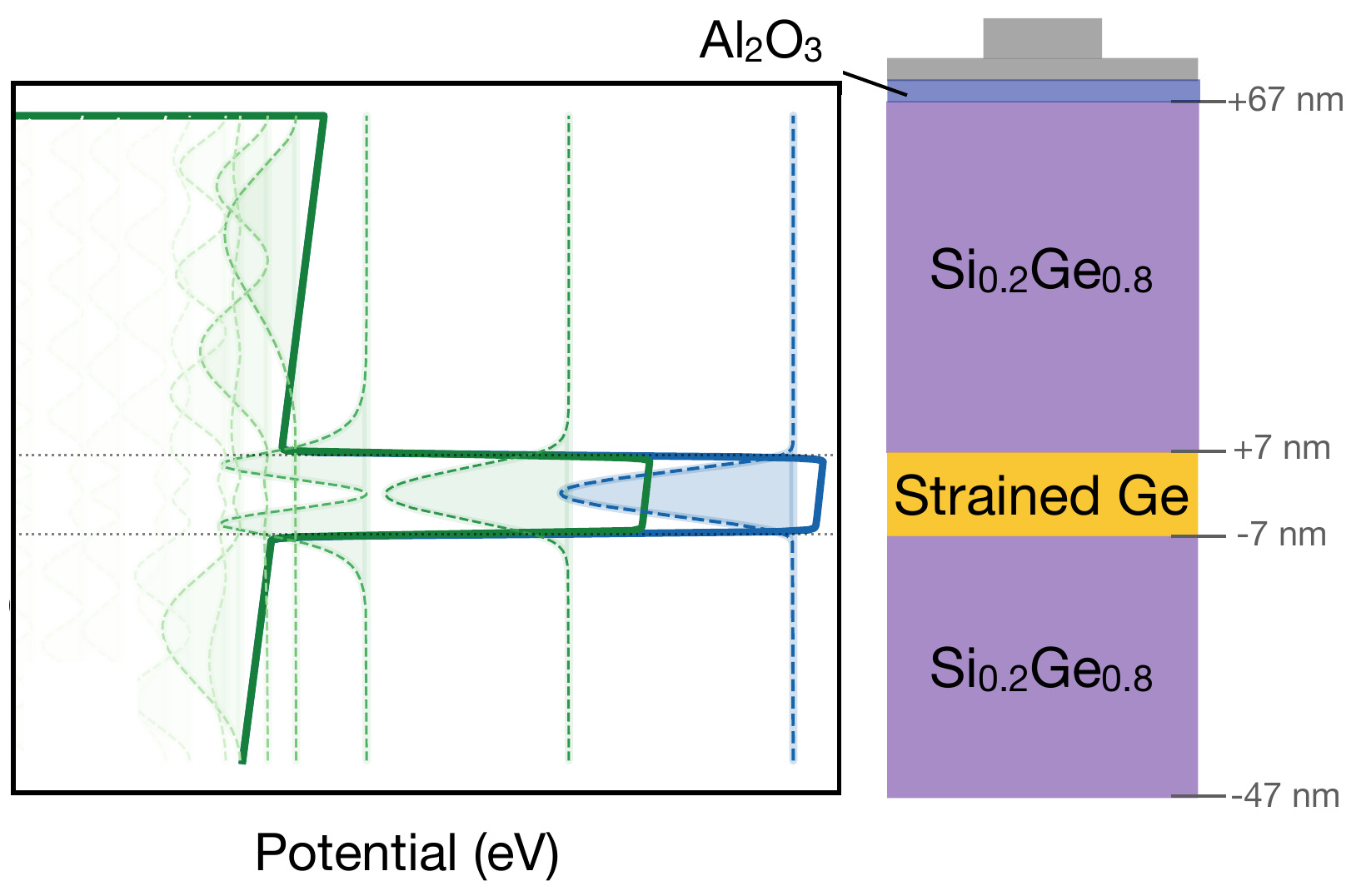}
  \caption{\textbf{Ge/SiGe hole dot cross-section and band diagrams.} Here, the growth direction is along $[001]\equiv\hat z$; in-plane axes are $\hat x\!\parallel\![100]$, $\hat y\!\parallel\![010]$. Dashed ticks mark the Ge QW interfaces at $z=\pm\SI{7}{nm}$. The $\mathrm{HH}$ (blue) and $\mathrm{LH}$ (green) out-of-plane potentials are shown versus $z$, along with the corresponding out-of-plane squared envelopes (dashed). Envelopes are vertically offset by their eigenenergies for visibility. The gate field $E_z$ adds a linear tilt to the potentials. Color and axis conventions are used consistently throughout.}
\label{fig:device_sidecut}
\end{figure}

We model a compressively strained Ge quantum well (QW), with thickness $L_z=\SI{14}{nm}$, embedded between relaxed SiGe barriers (see Fig.~\ref{fig:device_sidecut})~\cite{shimura2024}. Holes are electrostatically accumulated by a metal gate across an insulating oxide. The gate induces an average vertical electric field $E_z$ that both accumulates holes and tilts the valence band edges.

As introduced in Sec.~\ref{sec:spin_qubit_model}, the valence-band states are described within a multiband $\mathbf{k}\!\cdot\!\mathbf{p}$ framework that includes both Luttinger--Kohn and Bir--Pikus terms. In a compressively strained Ge QW grown along the $[001]$ direction, the biaxial strain imposed by the relaxed SiGe barriers (due to mismatch of lattice constants) lifts the HH–LH degeneracy at $\mathbf{k}=\mathbf{0}$, placing the HH band above the LH band~\cite{Scappucci_2021}. This results in a finite HH–LH energy splitting $\Delta_{\mathrm{HL}}=E^{(\mathrm{HH}_0)}-E^{(\mathrm{LH}_0)}>0$.

Although the ground state at $\mathbf{k}=\mathbf{0}$ is predominantly HH-like, off-diagonal terms of the Luttinger--Kohn Hamiltonian couple HH and LH subbands in the presence of confinement, leading to a finite LH admixture in the lowest Kramers pair~\cite{Stano_2025}. We refer to the two-dimensional subspace spanned by the lowest Kramers pair as the HH-like doublet. Unless stated otherwise, all $g$-factors reported in the following refer to this HH-like doublet.

Exploiting the predominantly HH character of the ground state at $\mathbf{k}=\mathbf{0}$, we treat HH–LH mixing perturbatively. Within this approximation, the low-energy Hamiltonian can be expressed via a Schrieffer–Wolff transformation in the following basis
\begin{equation}
    \Psi^{(j)}_{0,n}(x,y,z) = \psi^{(j)}_0(x,y)\; \phi^{(j)}_n(z), \; j \in \{\mathrm{HH}, \mathrm{LH} \} \; .
\end{equation}
This perturbative ansatz is valid when the HH–LH subband splitting dominates over the off-diagonal Luttinger–Kohn couplings, as is typical in strongly confined quantum wells ($L_z \ll L_x, L_y$)~\cite{Winkler2003_Book,Terrazos_2021,Martinez2022_NonSeparableFields}. Within this framework, the in-plane orbitals $\psi^{(j)}_0(x,y)$ are obtained from an effective-mass Schrödinger equation for a two-dimensional harmonic potential~\cite{Wang2024_npjQI} (see App.~\ref{app:proc:lateral} for further details), while the out-of-plane envelopes $\phi^{(j)}_n(z)$ define the HH and LH subbands that set their quantized energy levels ~\cite{Wang2024_npjQI}. More precisely, the HH and LH subband envelopes $\phi^{(j)}_n(z)$ are the eigenstates of the effective one-dimensional Hamiltonian
\begin{equation}
\Big[-\frac{\hbar^2}{2m_{j,\perp}}\partial_z^2 + V^{(j)}_{\perp}(z;\mathbf{s},E_z)\Big] \phi^{(j)}_n(z)=E^{(j)}_{n}\,\phi^{(j)}_n(z),
\label{eq:Hz_formal}
\end{equation}
with $j\in\{\mathrm{HH},\mathrm{LH}\}$ and $m^{(j)}_\perp$ the corresponding out-of-plane effective masses.

We solve Eq.~\eqref{eq:Hz_formal} on a finite growth-direction domain
$z\in[L^{(-)}_{\mathrm{b.c.}},\,L^{(+)}_{\mathrm{b.c.}}]$ that extends into the Si$_{0.2}$Ge$_{0.8}$ buffers on both sides of the Ge well (see Table~\ref{tab:num-geom} in App.~\ref{app:num}). At the upper boundary $z=L^{(+)}_{\mathrm{b.c.}}$, we impose a hard-wall (Dirichlet) condition,
$\phi^{(j)}_n(L^{(+)}_{\mathrm{b.c.}})=0$, reflecting the strong confinement at the gate-side oxide interface, where the wave function envelope is strongly suppressed. At the lower boundary $z=L^{(-)}_{\mathrm{b.c.}}$, we also impose $\phi^{(j)}_n(L^{(-)}_{\mathrm{b.c.}})=0$, but this condition is used as a practical numerical truncation rather than a physical interface. We choose $L^{(-)}_{\mathrm{b.c.}}$ sufficiently far into the lower buffer such that the relevant low-energy quantities, in particular the extracted $g_{xx}$, are converged and do not change appreciably upon further increasing the domain size.

The effective out-of-plane potential, $V^{(j)}_{\perp}(z;\mathbf{s},E_z)$, in Eq.(\ref{eq:Hz_formal}) is defined as
\begin{equation}
V^{(j)}_{\perp}(z;\mathbf{s},E_z) \;=\; W^{(j)}(z;\mathbf{s}) \;-\; eE_z z \; ,
\label{eq:V_total_formal}
\end{equation}
where the linear term $-eE_z z$ represents the applied electric
field potential. The term
$W^{(j)}(z;\mathbf{s})$ is the material and band-dependent valence-band-edge profile entering Eq.~\eqref{eq:Hz_formal}: it combines (i) the strain-induced band-edge shifts obtained from the Bir--Pikus Hamiltonian \cite{Wang2024_npjQI}, (ii) the Ge/Si$_{0.2}$Ge$_{0.8}$ valence-band offset \cite{Wang2024_npjQI}, and (iii) the local modification of the band edges due to the Si concentration inside the Ge well~\cite{Terrazos_2021} (see \ref{app:proc} for further details). To emulate layered planar heterostructure growth, $W^{(j)}(z;\mathbf{s})$ is parametrized by an $n_{\text{seg}}$-component vector
$\mathbf{s}=(s_1,\dots,s_{n_{\text{seg}}})$ that parametrizes a smooth, piecewise-constant Si profile across the well thickness (with smoothing introduced at segment interfaces). In general, the number of segments $n_{\text{seg}}$ is set by dividing the well thickness $L_z$ by the smallest composition-gradient length scale that can be resolved and meaningfully controlled in the intended fabrication process. Different fabrication platforms may naturally lead to different segment sizes and hence different values of $n_{\text{seg}}$. In the present work, we take this process-dependent length scale to be \SI{2}{nm}, which yields $n_{\text{seg}}=L_z/\SI{2}{nm}=7$ for the nominal $L_z=\SI{14}{nm}$ well.

The explicit construction of $W^{(j)}(z;\mathbf{s})$ and its strain and band-offset parameters are given in App.~\ref{app:proc:vertical}. We note that introducing Si inside the Ge well modifies the local strain and band profile, thereby providing a direct control parameter for tailoring $W^{(j)}(z;\mathbf{s})$ and the resulting subband spectrum.

Using the separable envelope ansatz and subband basis defined above, we assemble the multiband Hamiltonian, including Luttinger-Kohn, Bir-Pikus, and confinement terms, in the truncated product basis $\{\psi^{(j)}_0(x,y)\; \phi^{(j)}_n(z)\}$. The lowest Kramers doublet and its $g$-tensor are then obtained via the Schrieffer–Wolff transformation and projection procedure described in Sec.~\ref{sec:spin_qubit_model} (see Eqs.~\eqref{eq:H_B_expand}--\eqref{eq:Heff_g_tensor_def})~\cite{Wang2024_npjQI,Rimbach-Russ2025}. In the regime of large HH–LH splitting, for which HH–LH mixing remains perturbative, this multiband construction reduces to a compact perturbative expression for the in-plane response of a $[001]$-grown quantum well~\cite{Winkler2003_Book,Rimbach-Russ2025,Terrazos_2021}. For in-plane biaxial strain $\langle\epsilon_{xx}\rangle=\langle\epsilon_{yy}\rangle$ and negligible shear strain $\langle\epsilon_{xy}\rangle\simeq 0$, the leading confinement-induced correction to $g_{xx}$ yields
\begin{equation}
\begin{split}
\label{eq:gxx_formal}
g_{xx}(\mathbf{s}) & \approx  3q-
\frac{6}{m_0\Delta_{HL}(\mathbf{s})}\Big[\lambda(\mathbf{s})\langle p_x^2\rangle-\lambda'(\mathbf{s})\langle p_y^2\rangle\Big] \\
&\equiv 3q-\delta g^{\mathrm{conf}}_{xx}(\mathbf{s})\; ,
\end{split}
\end{equation}
where $\Delta_{\mathrm{HL}}=E^{(\mathrm{HH}_0)}-E^{(\mathrm{LH}_0)}>0$ is the gap between the lowest HH and LH subbands, and the coefficients $\lambda,\lambda'$ are functionals of the out-of-plane envelopes (see Eqs.~\eqref{lambda_eq_def}--\eqref{lambdap_eq_def} in App.~\ref{app:proc:gxx}). In our model, the separation of tuning parameters is explicit: the vertical potential $V^{(j)}_\perp(z;\mathbf{s},E_z)$ determines the HH–LH splitting $\Delta_{\mathrm{HL}}$ and the overlap-dependent prefactors in $\lambda$ and $\lambda'$, while the lateral confinement, parametrized by $L_x$ and $L_y$ for a harmonic dot, governs the kinetic moments $\langle p_x^2\rangle$ and $\langle p_y^2\rangle$ (see Eq.~\eqref{eq:gxx_formal}).

In a disk-shaped QD grown along the high-symmetry direction $[001]$, with in-plane biaxial strain and negligible shear strain, the symmetry of the system implies that the in-plane principal axes of the $g$-tensor coincide with the crystallographic directions $\hat{x}$ and $\hat{y}$. In this regime, the explicit expressions for $g_{xx}$ and $g_{yy}$ derived in Refs.~\cite{Rimbach-Russ2025, Martinez2022_NonSeparableFields, Michal2021_PRB_1DChannels} show that the in-plane $g$-tensor is diagonal and satisfies $g_{xx}=-g_{yy}$. Consequently, any tuning of the vertical confinement that reduces $g_{xx}$ will likewise reduce $g_{yy}$. Under these symmetric conditions, it is therefore sufficient to minimize $g_{xx}$ in the cost function. For completeness, we verified that $|g_{yy}|$ consistently follows the same trend as $g_{xx}$ across all optimized configurations (see App.~\ref{app:DeltaHL} for further details).

The optimization task is to determine a Si concentration profile $\mathbf{s}$ for which $V^{(j)}_\perp(z;\mathbf{s}, E_z)$ suppresses $g_{xx}$ while keeping the HH component of the HH-like doublet confined to the QW. For the nominal device, $L_z = \SI{14}{nm}$ and $L_x = L_y = \SI{40}{nm}$. The confinement-induced contribution $\delta g^{\mathrm{conf}}_{xx}$ can exceed $3q$, producing $g_{xx} < 0$ in Eq.~\eqref{eq:gxx_formal}; this is acceptable at the design stage, as it expands the accessible tuning range. Once fabrication fixes $\mathbf{s}$, the lateral confinement ($L_x$, $L_y$) provides a convenient control parameter for fine-tuning $g_{xx}$ to the target value, including $g_{xx} \simeq 0$.

\subsection{Optimization of the Si profile (CMA-ES)}
\label{sec:cmaes_opt}

To formulate our optimization procedure, we choose to discretize the quantum-well's Si profile in the out-of-plane direction, $z$, into $n_{\textit{seg}}=7$ equally spaced segments. Since the QW interfaces are located at $z=\pm L_z/2$, our analysis is restricted to the domain $z\in[-L_z/2,+L_z/2]$. Within this region, we allow the out-of-plane potential to be re-modulated through a piecewise-constant Si concentration profile, with the Si percentage permitted to vary independently in each segment. To avoid unphysical discontinuities in the potential, the transitions between adjacent segments are smoothed using a logistic-function interpolation (see App.~\ref{app:proc:vertical} for further details). The resulting out-of-plane potential is fully specified by the seven-component vector $\mathbf{s} = (s_1, \dots, s_7)$, with each $s_i$ controlling the Si fraction in the corresponding segment within the range $5\%-20\%$, i.e $s_i\in[5,20]\%$.

To emulate fabrication-limited precision in the Si concentration profile, candidate $\mathbf{s}$ vectors proposed by CMA-ES are rounded prior to cost-function evaluation. Rounding within the evaluation loop ensures that the optimizer identifies solutions robust to finite composition resolution, rather than unrealistically precise profiles. We note that the bounds $s_i\in[5,20]\%$ are chosen for both practical and physical reasons: $5\%$ reflects a fabrication-motivated lower limit for reliably defining a Si-containing segment of width $\SI{2}{nm}$ (process dependent), while $20\%$ matches the Si content of the Si$_{0.2}$Ge$_{0.8}$ buffers.

For the benchmark goal of achieving a gapless single-spin qubit encoding, the optimization task is to identify the vector $\mathbf{s}$ that minimizes $g_{xx}$, while ensuring proper confinement of the dominant HH component within the QW. This procedure can be summarized by the following steps. For each candidate $\mathbf{s}$ proposed by the optimizer, we:
\begin{enumerate}
    \item Assemble $V^{(j)}_\perp(z;\mathbf{s},E_z)$ for $j\in\{\mathrm{HH},\mathrm{LH}\}$.
    \item Solve the out-of-plane Schrödinger equation to obtain $\phi^{(\mathrm{HH})}_{0}(z)$ and $\{\phi^{(\mathrm{LH})}_{n}(z)\}_n$.
    \item Evaluate the in-plane HH ground-state orbital moments, $\langle p_{x}^2\rangle$ and $\langle p_{y}^2\rangle$, for a 2D harmonic oscillator.
    \item Compute $g_{xx}(\mathbf{s})$, using Eq.~\eqref{eq:gxx_formal}.
\end{enumerate}
We then minimize the cost-function

\begin{equation}
\label{eq:cost}
\mathcal{L}(\mathbf{s}) \;=\; g_{xx}(\mathbf{s}) \;+\; 3\,\max \;\!\bigl(0,\,P_{\mathrm{out}}(\mathbf{s})-0.4\bigr) \; ,
\end{equation}
where $P_{\mathrm{out}}=1-\int_{-L_z/2}^{L_z/2}|\phi^{(\mathrm{HH})}_{0}(z)|^2\,dz$ denotes the probability that the HH-component penetrates outside the nominal QW region. The threshold $P_{\mathrm{out}}=0.4$ is chosen as a practical confinement tolerance. It allows the HH envelope to extend slightly into the surrounding barriers, while still requiring most of the HH probability density to remain inside the nominal QW region. The second term in Eq.~\eqref{eq:cost} penalizes potential configurations that fail to ensure sufficient confinement of the HH-like doublet. We note that a small penetration of the out-of-plane HH wave function into the surrounding barriers is acceptable, provided that most of the wave function density remains confined within the QW. Additionally, while a larger penetration of HH wave functions into the barrier region naturally increases $g_{xx}$, scaling the penalty term by 3 further reinforces this effect. This rescaling was chosen to help the optimizer identify the desired range of solutions more efficiently.

In realistic experimental conditions, the out-of-plane electric field is not fixed to a single value but varies within an operating window
$E_z\in[E_{\min},E_{\max}]$. To ensure robust performance across this range, we adopt a conservative worst-case scenario design strategy. More precisely, the CMA-ES optimization is performed at the upper bound $E_{\max}$, yielding an optimized Si concentration profile $\mathbf{s}^*\equiv\mathbf{s}^*(E_{\max})$. This profile is then held fixed while the resulting $g_{xx}$ and confinement metric $P_{\mathrm{out}}$ are evaluated for all fields $E_z\le E_{\max}$ within the target window. In our simulations, this approach produces a design that maintains a small $g_{xx}$ and acceptable wave function confinement throughout $[E_{\min},E_{\max}]$. Fields exceeding $E_{\max}$, however, can drive the HH envelope into the oxide and thereby invalidate the optimized solution (see Sec.~\ref{sec:results} and App.~\ref{app:gvsE}).

In the CMA-ES, candidate solutions are sampled from a multi-variate Gaussian distribution, $\mathcal{N}(\boldsymbol{\mu},\sigma^2\mathbf{C})$, parametrized by a mean vector $\boldsymbol{\mu}$, a step-size $\sigma$ (determining the overall spread of the distribution), and a covariance matrix $\mathbf{C}$. We denote the set of all candidate solutions at each iteration as the population, $N_{\text{pop}}$, while a single candidate solution is referred to as an individual. By iteratively adapting the distribution parameters, the algorithm is designed to progressively bias the exploration of space towards more promising candidate solutions~\cite{hansen2006cma, nomura2024cmaessimplepractical}. We refer to App.~\ref{app:proc:opt} for further details on the CMA-ES applied to the cost function in Eq.~\eqref{eq:cost}.

\section{Results}
\label{sec:results}

\begin{figure}[t]
  \centering
  \includegraphics[width=\columnwidth]{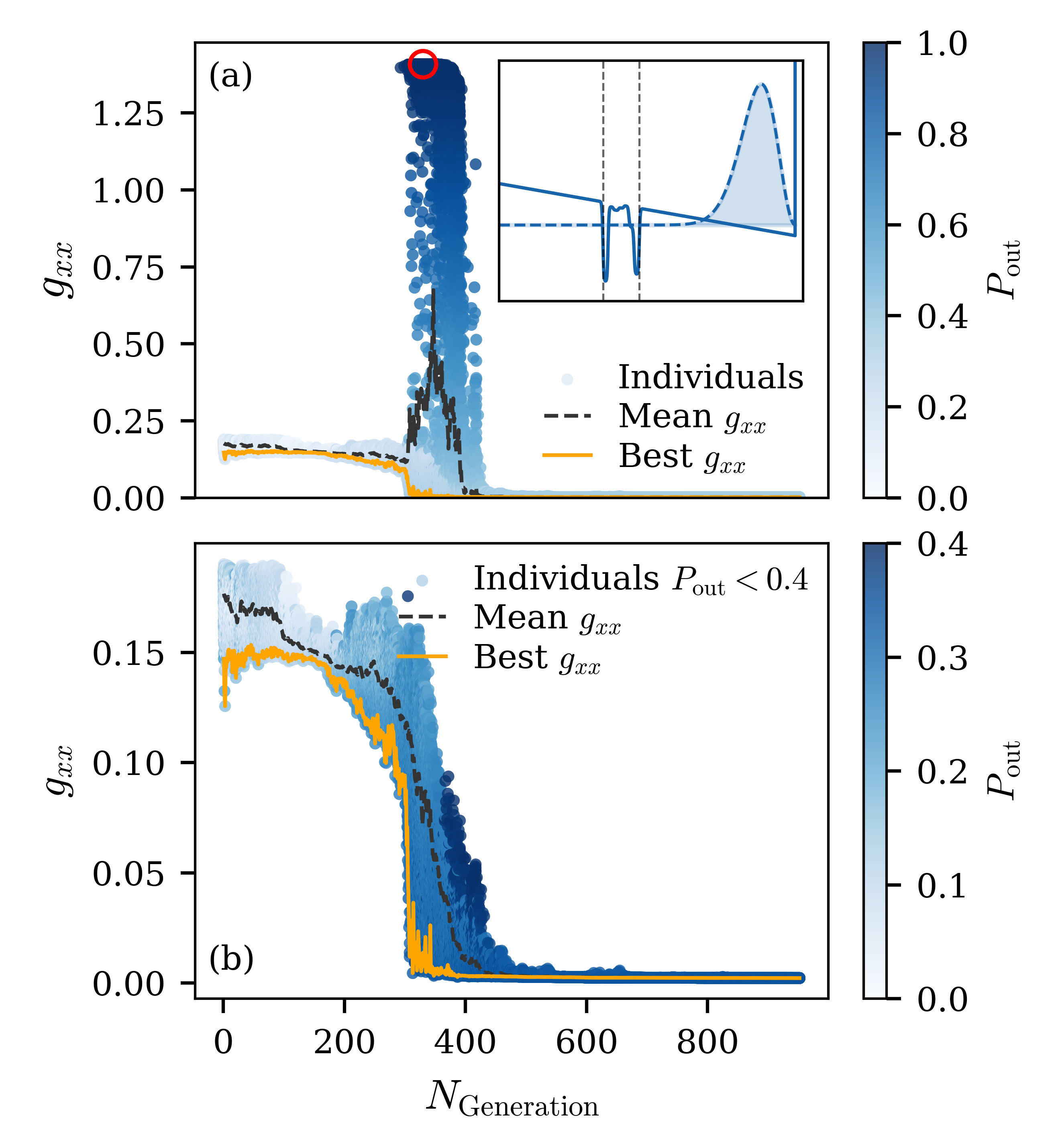}
\caption{\textbf{Convergence of in-plane response.} (a) Evolution of $g_{xx}$ across CMA-ES generations, $N_{\text{Generation}}$. Each point is a candidate Si profile. The different shades of blue indicate the probability, $P_{\mathrm{out}}$, that the HH-component of the HH-like doublet extends beyond the QW region. The inset figure displays the out-of-plane potential for one of the ill-confined solutions: the characteristic double-well profile has already formed, but still requires refinement to achieve the optimal $g_{xx}$ value. (b) The solid orange curve corresponds to the best-of-generation for $g_{xx}$, and the dashed gray curve is the population mean. The penalty term in $\mathcal{L}(\mathbf{s})$ (see Eq.~\eqref{eq:cost}) suppresses the probability of candidate solutions exiting the QW ($P_{\mathrm{out}}>0.4$), while still allowing $g_{xx}$ to be minimized.}
  \label{fig:gxx_vs_generation}
\end{figure}

The convergence of the CMA-ES towards the optimal Si profile $\mathbf{s}$  is illustrated in Fig.~\ref{fig:gxx_vs_generation}. In the early generations, the population explores a broad range of $\mathbf{s}$ vectors. As the optimization proceeds, the optimizer first identifies the overall potential shape required to achieve the optimization goals. This is illustrated by the intermediate generations, which harbor a significant amount of candidate solutions leading to HH wave functions that largely penetrate outside the QW (darker blue solutions in Fig.~\ref{fig:gxx_vs_generation} (a)). This occurs even though the resulting out-of-plane potential already entails the double-well-like shape characteristic of the optimum solution (see the inset in Fig.~\ref{fig:gxx_vs_generation} (a)). The subsequent stages of the optimization process focus on refining this potential shape to ensure the HH-component of the HH-like doublet is properly confined. This is supported by the observation that later generations predominantly explore candidate solutions that simultaneously minimize $g_{xx}$, while keeping the HH component well confined within the QW ($P_{\mathrm{out}}<0.4$).

In Fig.~\ref{fig:gxx_vs_generation} (b), the best-of-generation curve is shown to decrease monotonically. This indicates that once the optimizer finds the region of $\mathbf{s}$ vectors that both preserve the HH confinement and reduce $g_{xx}$, subsequent optimization updates consistently amplify those improvements. Meanwhile, the population mean converges towards the same value as the best-of-generation, signaling that nearly all individuals in the final generations produce similarly low $g_{xx}$ values and achieve good confinement. The convergence of the mean and best-of-generation curves indicates that the search distribution has collapsed around a single, well-defined optimum.

\begin{figure}[t]
  \centering
  \includegraphics[width=\columnwidth]{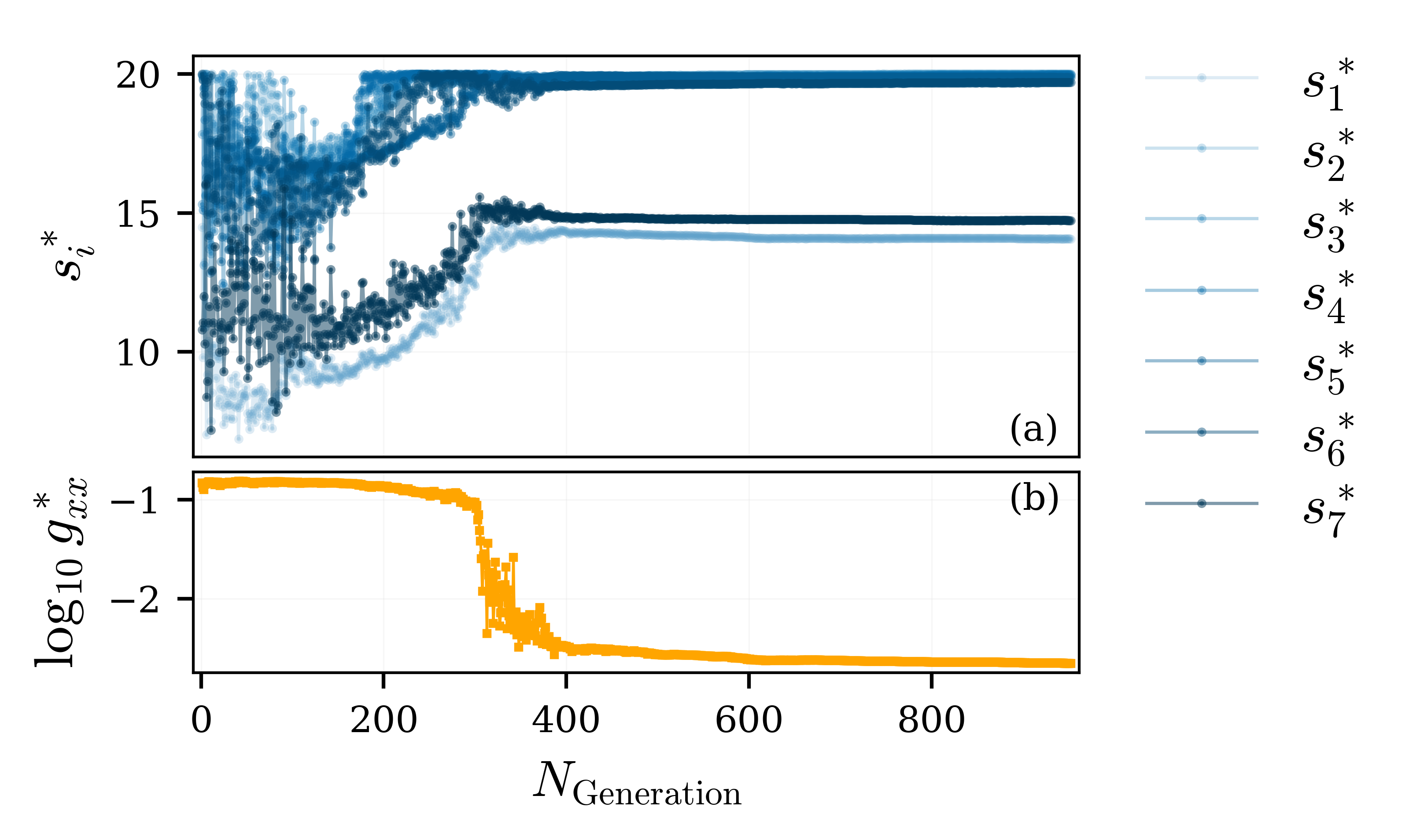}
  \caption{\textbf{Emergence of double-well pattern.} Evolution of the seven-segment Si vector $\mathbf{s}^*=(s_1^*,\dots,s_7^*)$ for the best candidate per generation (panel (a)) and the corresponding $\log_{10} g^*_{xx}$ (panel (b)). The search converges to the following Si distribution: low Si near the interfaces and nearly flat high-Si plateau in the center. This pattern minimizes $g_{xx}$.}
  \label{fig:segments_and_log}
\end{figure}

To illustrate how convergence towards the optimal solution is reflected in the Si concentration profile within the Ge well, we track the evolution of each of the seven segments over successive generations. This is shown in Fig.~\ref{fig:segments_and_log}, where panel (a) shows the evolution of the seven Si segments. Initially, the segment values fluctuate randomly across their full $[5,20]\%$ range. After a sufficient number of generations, the optimizer learns from feedback, and a clear pattern emerges: the Si concentration decreases near both interfaces, while the middle segments rise and flatten to form a uniform, high-Si plateau. This ``double-well'' pattern reshapes the vertical valence-band edges to minimize the in-plane Zeeman response. Panel (b) shows the corresponding evolution of $\log_{10} g_{xx}^*$, which drops monotonically from approximately $10^{-1}$ to $10^{-3}$, a difference of almost two orders of magnitude. The residual nonzero value is likely limited by the finite rounding and interpolation used in the Si-profile parametrization. This also leaves room for post-fabrication fine tuning through experimentally accessible control parameters. Once the optimized Si profile sets the baseline response, such fine-tuning can be used either to reach the target value $g_{xx}=0$ or to shift $g_{xx}$ slightly in a controlled way to enable gate operations.

\begin{figure}[t]
  \centering
  \includegraphics[width=\columnwidth]{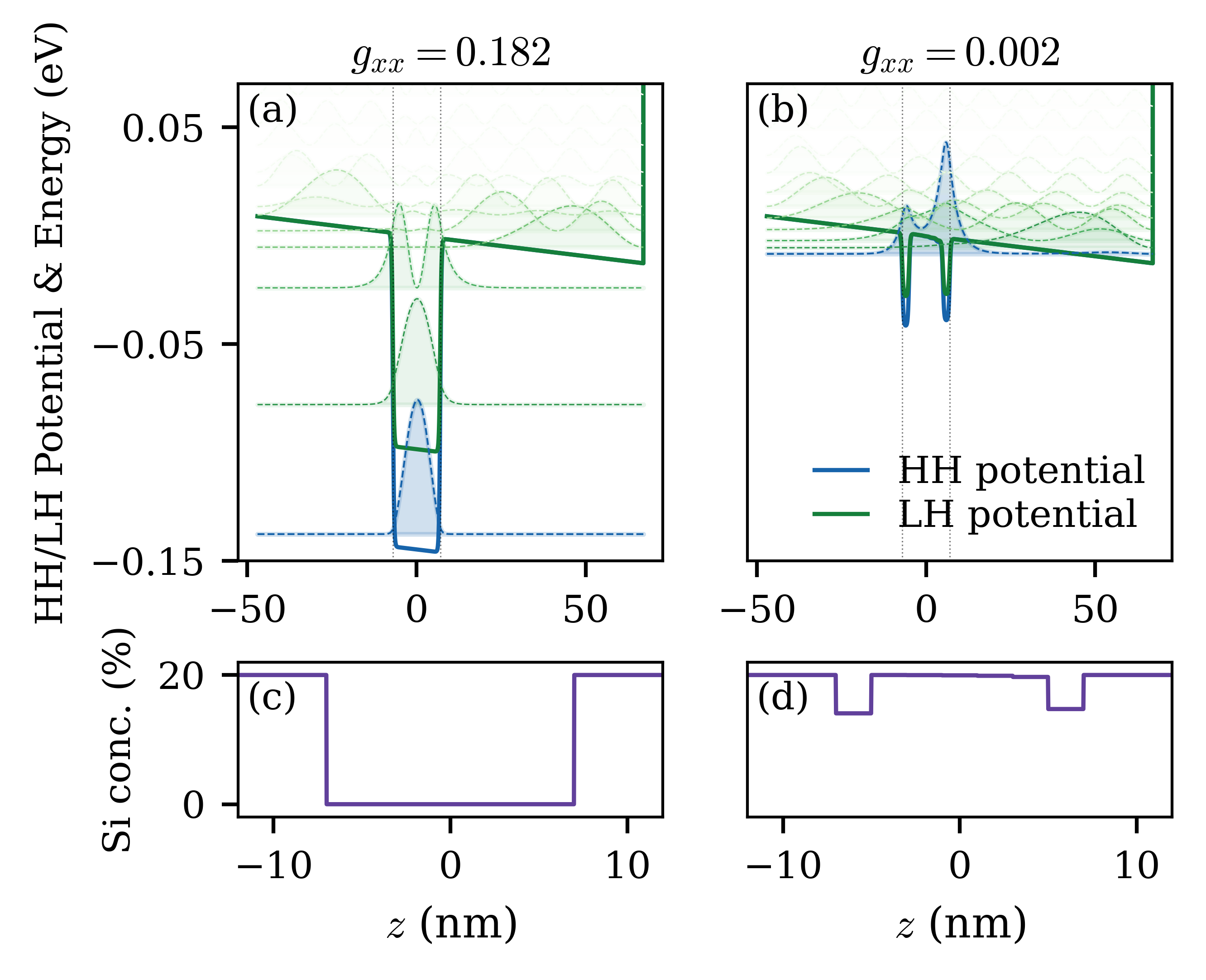}
  \caption{\textbf{Before/after vertical band edges and final improvement.}
The top panels display the out-of-plane potential (solid lines) and the squared envelope functions (dashed lines) for the HH (blue) and LH (green) subbands along the vertical coordinate (z), with potential energy in eV indicated on the vertical axis. The vertical dotted lines denote the Ge QW interfaces at $\pm \SI{7}{nm}$. The bottom panels provide the corresponding Si (\%) profiles within the well.}
  \label{fig:sigxx}
\end{figure}

A direct comparison between the optimized out-of-plane potential and the zero-Si-concentration baseline is shown in Fig.~\ref{fig:sigxx}, which displays both potentials together with their corresponding wave functions. In the zero–Si case (Fig.~\ref{fig:sigxx}, panel (a) and (c)), the $\mathrm{HH}$ wave function is centered and isolated from the interfaces. This configuration leads to a large $\mathrm{HH}$–$\mathrm{LH}$ gap. In the optimized structure (Fig.~\ref{fig:sigxx}, panel (b) and (d)), the two potential wells near the interfaces draw the HH envelope towards them. This increases the vertical overlap with $\mathrm{LH}$ subbands, thereby enhancing the confinement-induced $g$-tensor contribution, $\delta g^{\mathrm{conf}}_{xx}$, in Eq.~\eqref{eq:gxx_formal}. Despite this enhanced coupling, the $\mathrm{HH}$ probability density (dashed blue curve) remains well confined between the interface boundaries, showing that the optimization achieves strong mixing without delocalization. Thus, the double-well profile reshapes the valence-band edges so that the $\mathrm{HH}$ envelope couples optimally to the $\mathrm{LH}$ subbands while remaining confined. Quantitatively, this reshaping lowers $g_{xx}$ from $0.182$ to $0.002$, corresponding to two orders of magnitude suppression of the in-plane Zeeman response. We note that extending the range of Si concentration to $[5,30]\%$ yields the same optimum solution, with the plateau concentration still saturating near $\sim20\%$. This indicates that the plateau is set by the physics rather than the imposed bound: a buffer-matched (Si$_{0.2}$Ge$_{0.8}$-like) plateau maximizes HH--LH mixing and reduces the energy gap $\Delta_{\mathrm{HL}}$.

\begin{figure}[t]
  \centering
  \includegraphics[width=\columnwidth]{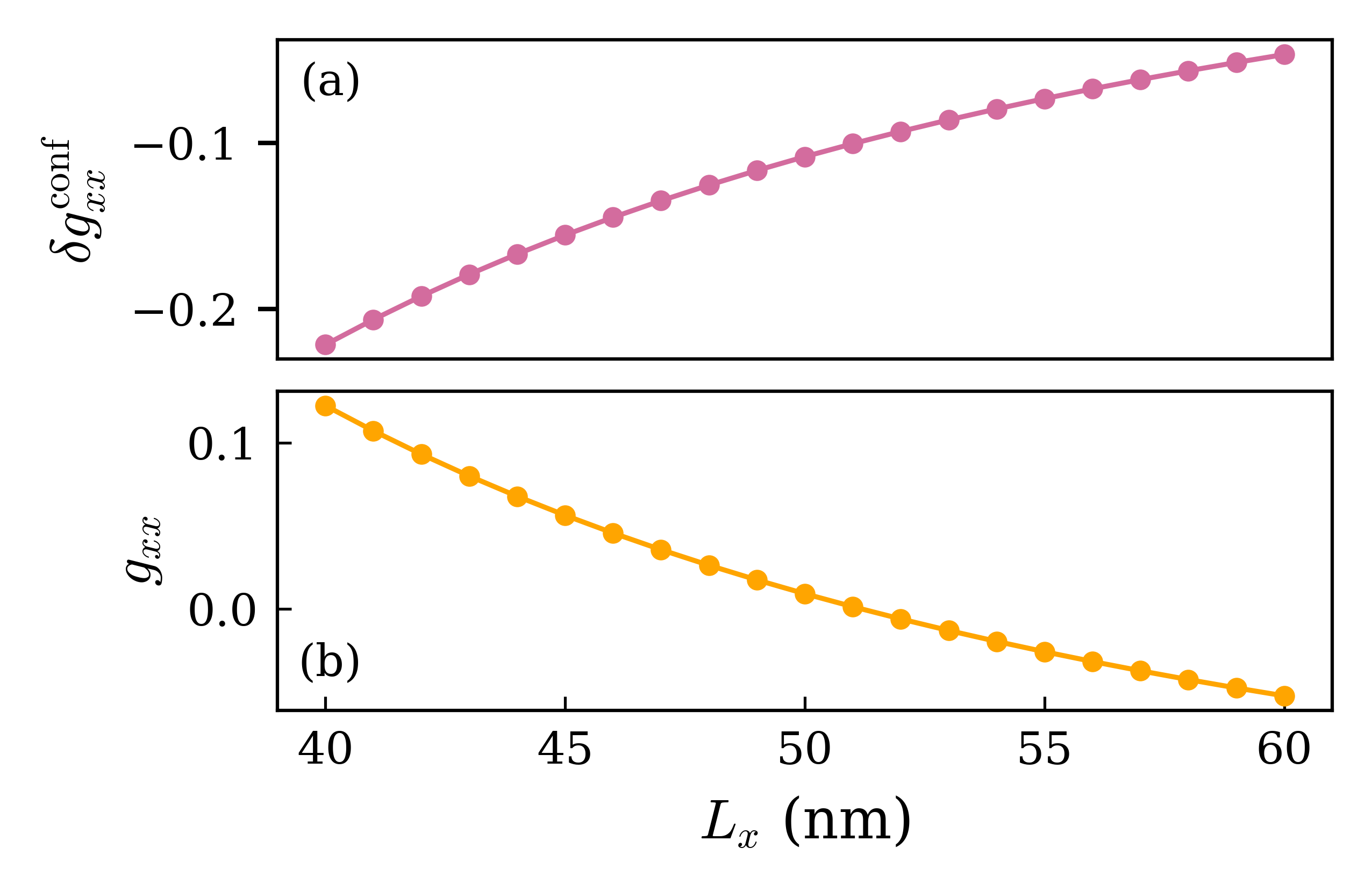}
  \caption{\textbf{Lateral tunability of the in-plane Zeeman response.}
  Confinement-induced correction $\delta g^{\text{conf}}_{xx}$ (top) and total in-plane response $g_{xx}$ (bottom) as a function of the lateral confinement length $L_x$, for a fixed vertical field $E_z=\SI{0.19}{MV/m}$ and the quantum-well thickness $L_z=\SI{14}{nm}$. The dot is isotropic at the idle operating point $L_x=L_y=\SI{40}{nm}$, where the optimized Si profile yields $g_{xx}\approx 0$. Varying $L_x$ away from this point changes the kinetic moment, thereby tuning the confinement-induced reduction $\delta g^{\text{conf}}_{xx}$ and shifting $g_{xx}$ away from the gapless condition. This demonstrates that small distortions of the lateral dot geometry can be used as a low-frequency control parameter for qubit operations within the gapless single-spin qubit encoding regime.}
  \label{fig:gxx_Lx}
\end{figure}

The achieved suppression of $g_{xx}$ already places a potential device into the gapless single-spin qubit regime~\cite{Rimbach-Russ2025}. Achieving $g_{xx}=0$ exactly is not required at the optimization stage, as subsequent adjustments of the lateral confinement parameters $(L_x,L_y)$ provide an effective means to further tune $g_{xx}$ towards the desired near-zero regime. This tunability is illustrated in Fig.~\ref{fig:gxx_Lx}, where varying $L_x$ at fixed $L_y$ induces a controlled shift of the confinement-induced contribution $\delta g^{\mathrm{conf}}_{xx}$, thereby enabling calibration of $g_{xx}$. Operationally, this tunability enables baseband qubit control within the gapless single-spin qubit encoding proposed in Ref.~\cite{Rimbach-Russ2025}: gate voltages that slightly induce dot anisotropy ($L_x\neq L_y$) move the system away from the degeneracy condition, $g_{xx}=0$, producing a finite and electrically tunable Zeeman splitting. In this sense, the in-plane dot geometry acts as an additional low-frequency control parameter, complementary to vertical-field tuning, providing a straightforward and robust route to electrically driven qubit operations.

\begin{figure}[t]
  \centering
\includegraphics[width=\columnwidth]{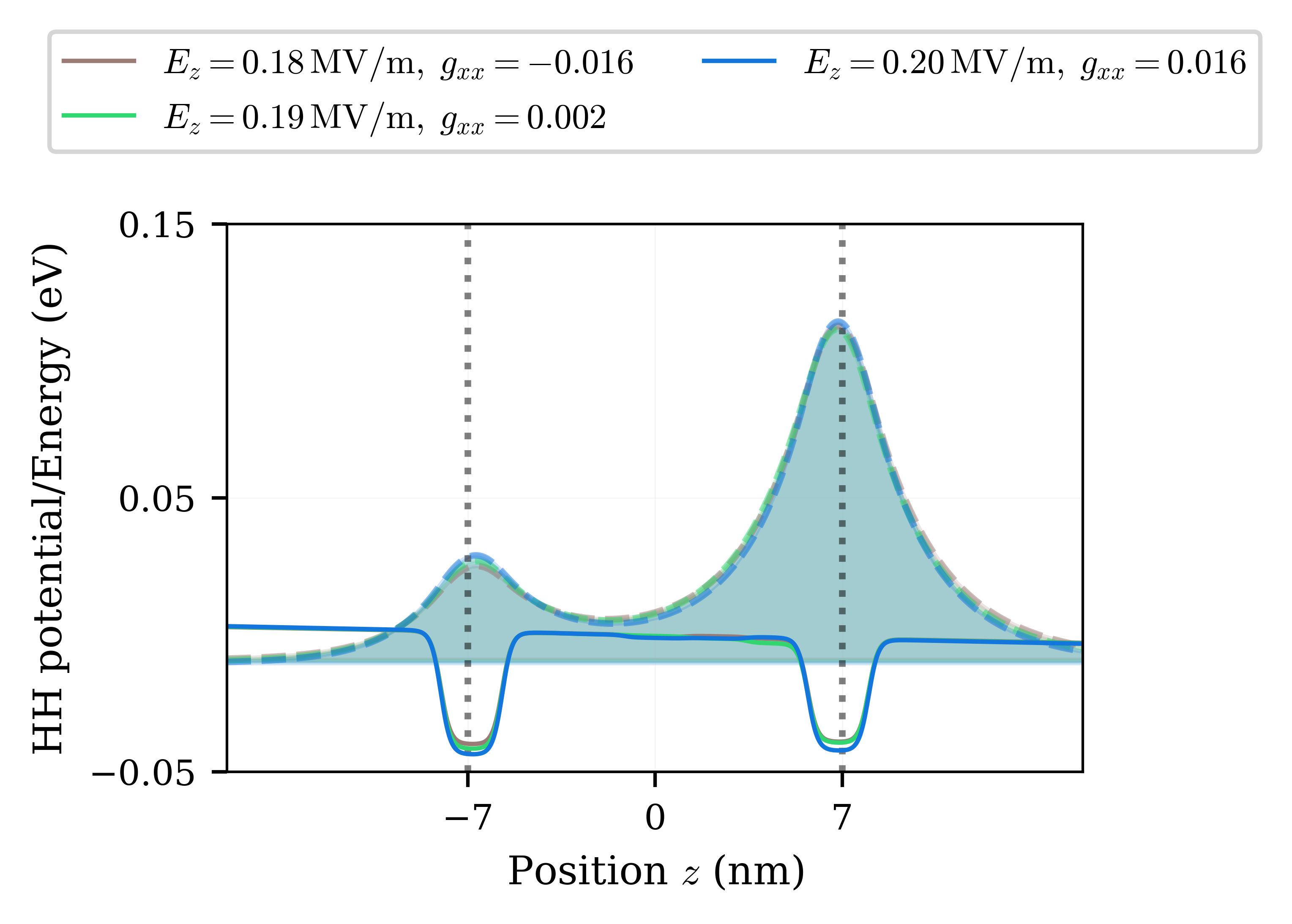}
  
\caption{\textbf{Robustness to out-of-plane electrical-field fluctuations.} Potential $V^{(\mathrm{HH})}_\perp(z)$ (solid) and $\mathrm{HH}$ ground-state probability density $|\phi^{(\mathrm{HH})}_{0}(z)|^2$ (dashed) for $E_z=\SI{0.18}{MV/m}$, $\SI{0.19}{MV/m}$, and $\SI{0.20}{MV/m}$. The optimized profile maintains confinement and low $|g_{xx}|$ across this range, demonstrating tolerance to gate-field quasi-static fluctuations.}
  \label{fig:field_robustness}
\end{figure}

The sensitivity to small quasi-static variations of the out-of-plane electric field (a proxy for device-to-device electrostatic variability) is examined in Fig.~\ref{fig:field_robustness}, which shows the optimized out-of-plane potential for three representative field strengths. Increasing or decreasing $E_z$ by $\pm\SI{0.01}{MV/m}$ changes the overall tilt of the potential, but does not affect the localization of the HH component. The extracted in-plane responses are $g_{xx}=-0.016$, $0.002$, and $0.016$ for the three field values, demonstrating that $|g_{xx}|$ remains small. This robustness means that the $g_{xx}$ suppression remains effective even when fluctuations of the electric field are present. Such resilience is crucial for reproducible and practical qubit operations, where quasi-static electrostatic offsets on the scale of a few kV/m are unavoidable~\cite{Valvo2025ElectricallyTuneable}. For larger offsets, $g_{xx}$ can deviate further from zero even though the wave function remains well-confined. In that case, $(L_x,L_y)$ provide a post-fabrication recalibration parameter to tune $g_{xx}$ back to the target value.

\section{Adaptable Simulation Parameters}
\label{sec:knobs}
The optimization framework introduced in this work includes several parameters that can be readily modified to suit different use cases. In this section, we outline these adjustable parameters and group them into three main categories: barrier parameters, quantum-well parameters, and the input optimization parameters.

Barrier parameters specify the characteristics of the environment in which the Ge quantum well is embedded. In our model, these include the Si concentration in the relaxed Si$_{0.2}$Ge$_{0.8}$ barriers and the thickness of the upper buffer/oxide region. Hence, they can be easily adapted to reflect different heterostructures.

The quantum-well parameters set the baseline confinement potential and parametrize the out-of-plane modulation caused by adding Si within the well. Accordingly, they include not only the quantum-well thickness $L_z$ and the in-plane confinement lengths $L_x$ and $L_y$, but also the parameters that describe the piecewise potential arising from the Si concentration profile. These comprise the number of segments $n_{\mathrm{seg}}$ used to construct the piecewise potential (chosen here as $n_{\mathrm{seg}}=7$; see App.~\ref{app:num:bandedges}), the logistic smoothing length $\ell$ that determines the smoothness between segment transitions, and the allowable Si concentration range $s_i\in[s_{\min},s_{\max}]$ for each segment. We note that the value of $\ell$ should reflect the physically achievable gradient of the Si concentration in the heterostructure. Extremely sharp transitions (small $\ell$) are unrealistic, while larger segment sizes allow for more gradual smoothing. Thus, these parameters can be adjusted to accommodate different fabrication constraints, but remain fixed throughout the optimization loop.

The scalar cost function $\mathcal{L}(\mathbf{s})$ can also be treated as an adjustable input, and in this work it is chosen to drive the HH-like doublet towards the gapless single-spin encoding operating point ($g_{xx},g_{yy}\to 0$)~\cite{Rimbach-Russ2025}. Other qubit design objectives can be implemented by modifying the cost function. For example, to engineer a more isotropic hole-spin qubit~\cite{MauRodMena2025}, one could replace $g_{xx}$ with a weighted combination such as $Bg_{zz}-Ag_{xx}$, shifting the optimization towards a more isotropic $g$-tensor. The penalty term can similarly be adapted to impose additional constraints on the target solutions. Thus, while our analysis focuses on a specific optimization objective, the methodology is broadly applicable and can be tailored to other $g$-tensor design targets.

In addition to the user-defined parameters discussed above, certain fixed parameters, primarily those affecting the numerical convergence of the solvers, may require adjustment when applying the method in different contexts. These convergence-related parameters are discussed in detail in App.~\ref{app:conv_parameters}.

\section{Conclusion and Discussion}
\label{sec:discussion}

Currently, anisotropy and device-to-device variability in semiconductor quantum devices are major obstacles to scaling these systems to larger numbers of qubits. In hole-spin qubits in particular, precise control over the anisotropic spin properties of valence-band hole-spin states is crucial for reaching operational sweet spots. In this work, we presented an optimization strategy for engineering the in-plane $g$-tensor of hole spin qubits through targeted incorporation of silicon (Si) in a strained planar germanium quantum well. By combining an effective low-energy model with CMA-ES optimization, we identify the Si concentration profile that minimizes the in-plane $g$-tensor components of the HH-like doublet. Moreover, the resulting optimized in-plane Zeeman response remains robust under realistic quasi-static electric-field fluctuations of the order of $\pm\SI{0.01}{MV/m}$.

Overall, our findings present a stepping stone towards improved uniformity in germanium quantum devices. We deliver a proof-of-principle simulation indicating that targeted Si doping along the growth direction constitutes a viable approach to controlling the highly anisotropic spin properties of hole-spin states. This strategy has the potential to support reproducible, electrically driven qubit operations. Because the optimization framework is general, it can be applied to tailor other qubit properties, such as increasing valley splitting in Si quantum dots~\cite{Yang_2013} or modulating the spin--orbit coupling strength in hole-based systems~\cite{Xiong_2021}.

Although we have focused exclusively on a single-parameter objective function, a natural extension of this work is to incorporate a multi-objective formulation that accounts for various trade-offs, such as simultaneously controlling the $g$-tensor anisotropy and the Rabi frequency, within the same CMA-ES optimization loop.

During the preparation of this manuscript, we became aware of a related independent study employing machine learning to improve quantum-well properties~\cite{VecRosScap2026}.

\section{Acknowledgments}

This research was sponsored by the Army Research Office and was accomplished under Award Number: W911NF-23-1-0110 and partly by the EU through H2024 QLSI2.  The views and conclusions contained in this document are those of the authors and should not be interpreted as representing the official policies, either expressed or implied, of the Army Research Office or the U.S. Government. The U.S. Government is authorized to reproduce and distribute reprints for Government purposes notwithstanding any copyright notation herein.
This publication is part of the ’Quantum Inspire – the Dutch Quantum Computer in the Cloud’ project (with project number [NWA.1292.19.194]) of the NWA
research program ’Research on Routes by Consortia (ORC)’, which is funded by the Netherlands Organization for Scientific Research (NWO). This work is part of the project Engineered
Topological Quantum Networks (Project No.VI.Veni.212.278) of the research program NWO
Talent Programme Veni Science domain 2021 which is financed by the Dutch Research Council (NWO). M.R.-R. and E.V. additionally acknowledge support from the Dutch Research Council (NWO) under Award Number Vidi TTW 22204.

\section{Data and code availability}
The code and data used in this work are openly available at \url{https://gitlab.com/QMAI/papers/germaniumoptimization} and  \cite{shojaei_2026_19692317}.

\appendix
\section{Optimization scheme overview}    
\label{app:proc}
In this appendix, we provide additional details on the workflow underlying our optimization scheme. For clarity, we begin by recalling its three main steps:
\begin{enumerate}
    \item  We start by solving the out-of-plane (vertical) and in-plane (lateral) effective mass Schrödinger equations.
    \item  We then proceed to evaluate $g_{xx}$ via the Schrieffer-Wolff transformation of the Luttinger–Kohn model, in the presence of an applied in-plane magnetic field and a confinement potential. 
    \item Finally, using CMA-ES, we optimize the segmented Si profile concentration inside the Ge QW.
\end{enumerate}

Numerical settings, such as spatial discretization grids, boundary conditions, potential smoothing settings, and material constants, are given in App.~\ref{app:num}.

\subsection{Modeling the out-of-plane subbands}
\label{app:proc:vertical}
We consider a compressively strained Ge QW of thickness $L_z$, grown along $[001]\!\equiv\!\hat z$ and embedded between relaxed Si$_{0.2}$Ge$_{0.8}$ barriers with interfaces at $z_s=-L_z/2$ and $z_e=+L_z/2$. Inside the well, a piecewise-constant Si profile, $\mathbf{s}$, reshapes the out-of-plane confinement potential and, consequently, the $\mathrm{HH}$/$\mathrm{LH}$ valence-band edges.

For band $j\in\{\mathrm{HH},\mathrm{LH}\}$, the total vertical (out-of-plane) potential is defined as:
\begin{equation}
V^{(j)}_\perp(z;\mathbf{s})=W^{(j)}(z;\mathbf{s})-eE_z z \;,
\end{equation}
where $E_z$ is the gate-produced vertical electric field, and $W^{(j)}(z;\mathbf{s})$ acts as an effective potential that takes into account both the effects of strain and band offset, as well as the Si concentration inside the Ge well. Namely, $W^{(j)}(z;\mathbf{s})$ is defined as follows~\cite{Wang2024_npjQI}:

\begin{equation}
W^{(j)}(z;\mathbf{s})=
\begin{cases}
U_0^{(j)}-\dfrac{U_0^{(j)}}{c_{\mathrm{Si}}}\,s(z) & \; , \; z\in[z_s,z_e]\\[5pt]
0 \; , & \text{otherwise}
\end{cases}
\label{eq:Upiece-app}
\;\;\; ,
\end{equation}

where $U_0^{(j)}$ denotes the strain-shifted valence-band offset between Ge and Si$_{0.2}$Ge$_{0.8}$ layers, for either heavy-hole ($j=$HH) or light-hole ($j$=LH) states (see App.~\ref{app:num:bandedges} for further details on the effect of strain on the valence-band offsets). The linear character of Eq.~\eqref{eq:Upiece-app} ensures the interpolation between the two extreme limits: the limit $s=0 \%$, which reproduces the pure-Ge edge $U_0^{(j)}$, and the limit $s=c_{\mathrm{Si}}=20 \%$, which raises the edge to the barrier reference ($0$ outside the well). 

\begin{figure}[t]
  \centering
  \includegraphics[width=\columnwidth]{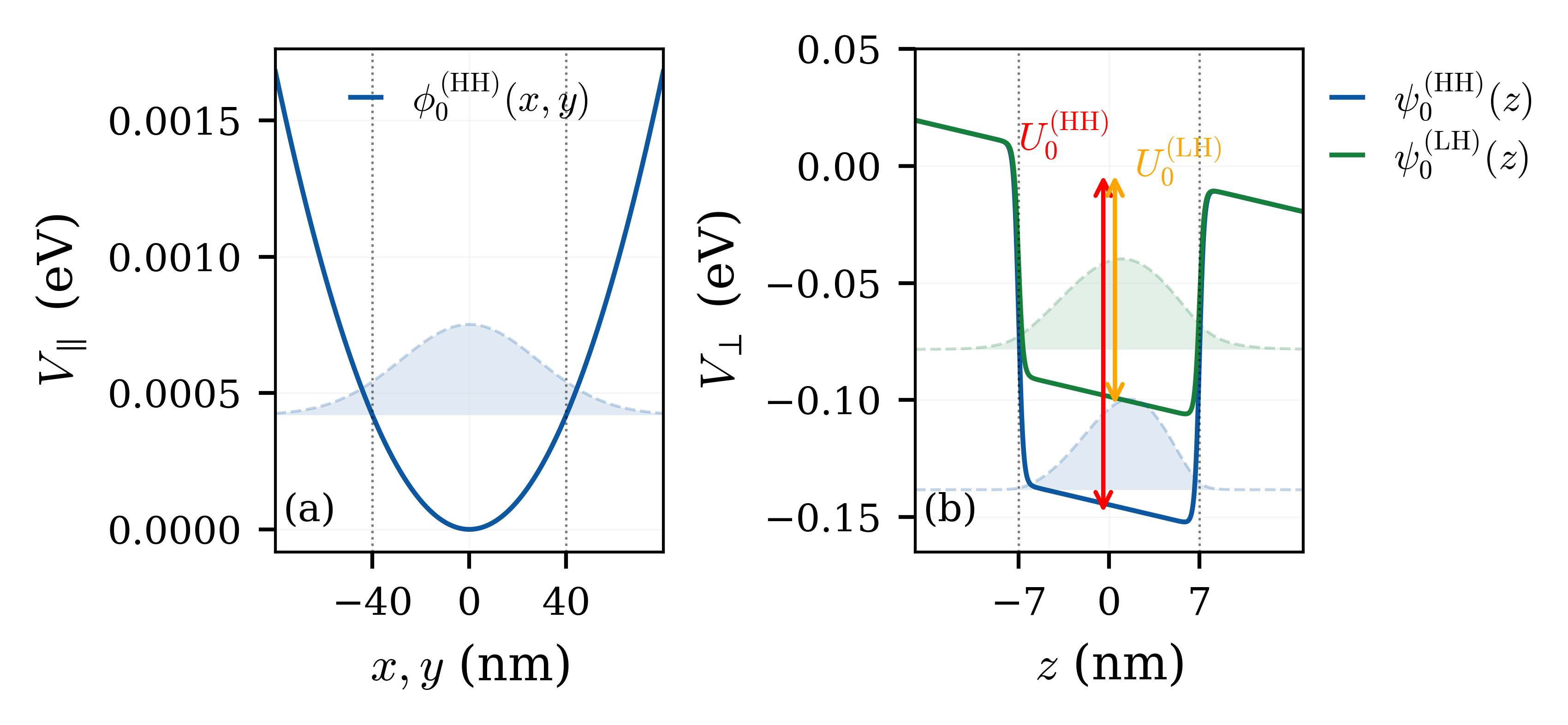}
  \caption{\textbf{Potential landscape.} (a) In-plane harmonic oscillator potential ($L_{x,y}=\SI{40}{nm}$) for the $\mathrm{HH}$ ground state orbital $\psi^{(HH)}_0(x,y)$. (b) Out-of-plane potentials $V^{\mathrm{HH}}_\perp(z;\mathbf{0})$ (blue) and $V^{\mathrm{LH}}_\perp(z;\mathbf{0})$ (green) subbands. Dashed lines indicate the strain-shifted reference edges $U^{(\mathrm{HH})}_0$, $U^{(\mathrm{LH})}_0$.}
  \label{fig:potentials}
\end{figure}

However, rather than working with Eq.~\eqref{eq:Upiece-app} directly, we employ a smoothed version of the potential to remove unphysical kinks. Specifically, each discontinuous transition in Eq.~\eqref{eq:Upiece-app} is replaced by a smooth logistic-function transition \cite{Beznasyuk_2020, Han2017_thesis}. Hence, instead of Eq.~\eqref{eq:Upiece-app}, we employ in simulations the following smoothed version of the potential $W^{(j)}(z;\mathbf{s})$: 

\begin{equation}
W^{(j)}(z;\mathbf{s})
 \to U^{(j)}_0 + \sum_m \Delta_m\,
   \frac{1}{1 + \exp\!\bigl[-(z - z_m)/\ell\bigr]} \; .
\label{eq:logistic}
\end{equation}
Here, $\ell$ controls the steepness of the transition curve, $\Delta_m$ is determined by the height of the two endpoints being connected, and $z_m$ specifies the midpoint of the connecting curve. For this profile, the valence-band edge changes from $10\%$ to $90\%$ of the step height over a distance $\Delta z_{10\text{--}90}\approx 4.4\,\ell$. In the simulations, we chose $\ell = \SI{0.20}{nm}$ (Table~\ref{tab:num-geom}), corresponding to
$\Delta z_{10\text{--}90} \approx \SI{0.88}{nm}$, i.e.\ a relatively sharp, but a numerically smooth transition. Smoothing is applied identically to $\mathrm{HH}$ and $\mathrm{LH}$ band edges, using their respective $U^{(j)}_0$ values (see Fig.~\ref{fig:potentials}). This yields the continuous potential $W^{(j)}$ entering in Eq.~\eqref{eq:V_total_formal} in the main text.

We then proceed to obtain the out-of-plane wave functions, $\phi^{(j)}_0(z)$. These are the solutions of the one-dimensional effective-mass Schrödinger equation:
\begin{equation}
\Big[-\frac{\hbar^2}{2m^{(j)}_{\perp}}\partial_z^2 + V^{(j)}_\perp(z;\mathbf{s})\Big]\phi^{(j)}_n(z)=E^{(j)}_{n}\phi^{(j)}_n(z) \; ,
\label{eq:Sheq}
\end{equation}
with
\begin{equation}
   m^{(HH)}_{\perp} = \frac{m_0}{\gamma_1-2\gamma_2} \; ,
\qquad
m^{(LH)}_{\perp} = \frac{m_0}{\gamma_1+2\gamma_2}  \; .
\end{equation}
From all HH subbands, we retain only the lowest energy subband,  $\phi^{(\mathrm{HH})}_0$, which provides the dominant component of the HH-like doublet. For LH subbands, we retain a larger sub-set of subbands,  $\{\phi^{(\mathrm{LH})}_n\}$. This sub-set of subbands contributes via virtual admixture to the final HH-like doublet. 

The confinement of the dominant $\mathrm{HH}$ component is controlled by the probability of penetrating outside the QW:
\begin{equation}
P_{\mathrm{out}} = 1 - \int_{z_s}^{z_e} |\phi^{\mathrm{HH}}_0(z)|^2\,dz \;.
\end{equation}
Larger values of $P_{\mathrm{out}}$ indicate that a bigger portion of the $\mathrm{HH}$ wave function density extends outside the QW, signaling poor confinement. In order to obtain a HH wave function that is well confined within the QW, we tracked $P_{\mathrm{out}}$ during simulations and regarded only those solutions satisfying $P_{\mathrm{out}}<0.4$ as fulfilling the confinement requirement.

\subsection{Modeling the in-plane wave functions} 
\label{app:proc:lateral}

We model the lateral confinement by a 2D harmonic potential, given by: 
\begin{equation}
V_\parallel^{(j)}(x,y) = \tfrac{1}{2}\,m^{(j)}_{\parallel}\!\left((\omega^{(j)}_x)^2 x^2+(\omega^{(j)}_y)^2 y^2\right) \; ,
\label{HO_potential}
\end{equation}
with
\begin{equation}
\omega^{(j)}_x=\frac{\hbar}{m^{(j)}_{\parallel}(L^{(j)}_x)^2}\; , \qquad
\omega^{(j)}_y=\frac{\hbar}{m^{(j)}_{\parallel}(L^{(j)}_y)^2} \; ,
\end{equation}
\begin{equation}
    m^{(\mathrm{HH})}_\parallel = \frac{m_0}{\gamma_1+\gamma_2} \; , \qquad
m^{(\mathrm{LH})}_\parallel = \frac{m_0}{\gamma_1-\gamma_2} \; ,
\end{equation}
\begin{equation}
    L_{x,y}^{(\mathrm{LH})} \;=\; L_{x,y}^{(\mathrm{HH})}
\left(\frac{\gamma_{1}-\gamma_{2}}{\gamma_{1}+\gamma_{2}}\right)^{\!1/4}.
\end{equation}
Note that $L_x^{(j)}$ and $L_y^{(j)}$ denote the harmonic oscillator characteristic lengths for each band. Since the harmonic potential in Eq.~\eqref{HO_potential} is a separable potential, the in-plane orbital wave functions are the product of two one-dimensional harmonic oscillator eigenstates, namely,  $\psi^{(j)}_0(x,y)=\psi^{(j)}_0(x) \; \psi^{(j)}_0(y)$, where

\begin{equation}
\begin{split}
\psi^{(j)}_{n}(x)=&\frac{(m_{j,\parallel}\omega_{x,j}/\hbar\pi)^{1/4}}{\sqrt{2^n n!}}\,
H_n\!\Big(\sqrt{\tfrac{m_{j,\parallel}\omega_{x,j}}{\hbar}}\,x\Big)\, \\
&\times e^{-m_{j,\parallel}\omega_{x,j}x^2/2\hbar} \; ,
\end{split}
\end{equation}

\begin{equation}
\begin{split}
\psi^{(j)}_{n}(y)&=\frac{(m_{j,\parallel}\omega_{y,j}/\hbar\pi)^{1/4}}{\sqrt{2^n n!}}\,
H_n\!\Big(\sqrt{\tfrac{m_{j,\parallel}\omega_{y,j}}{\hbar}}\,y\Big)\, \\
& \times e^{-m_{j,\parallel}\omega_{y,j}y^2/2\hbar} \; ,
\end{split}
\end{equation}

where $H_n$ is the $n$-th Hermite polynomial (see Fig.~\ref{fig:potentials} for a depiction of the $\mathrm{HH}$ ground state in-plane wave function).

We note that the in-plane orbitals contribute to the in-plane Zeeman response through the kinetic moments $\langle p_x^2\rangle \; \text{and} \;\langle p_y^2\rangle$. In particular, the in-plane $\mathrm{HH}$ ground state determines the orbital moments $\langle p_x^2\rangle$ and $\langle p_y^2\rangle$, which scale as $\hbar^2/L_{x,y}^2$ and contribute to the calculation of $g_{xx}$.

\subsection{Modeling the g-tensor response in hole qubits}
\label{app:proc:gxx}
The kinetic energy of holes in planar germanium is well-described by the $4 \times 4$ Luttinger--Kohn (LK) Hamiltonian defined in the $J=3/2$ valence band manifold, near $\mathbf{k}=\mathbf{0}$~\cite{Wang2024_npjQI}. In the absence of an applied magnetic field, the full Hamiltonian describing the holes is given by the LK Hamiltonian and the confinement potential $V(z;\mathbf{s})$, which encodes the QW profile and strain via Bir–Pikus formalism~\cite{Winkler2003_Book}.
Applying a magnetic field $\mathbf{B}$ changes the full Hamiltonian by the addition of the Zeeman interaction term, $H_{\text{Z}}$, and by the substitution of the momentum operator with the generalized momentum $\mathbf{k}\to \mathbf{k} + \frac{e}{\hbar}\mathbf{A}$, where we adopt the gauge choice: $\quad\mathbf{A} = (B_y z)\, \hat{\mathbf{x}} + (- B_x z)\, \hat{\mathbf{y}}$ \cite{Sarkar2023_PRB}. The Hamiltonian of the system is then given as:
\begin{equation}
H
=
\underbrace{H_{LK}(\mathbf{k}) + V(z;\mathbf{s})}_{H_0}
+ \underbrace{
H_{\mathrm{Z}}
+
\overbrace{
\left (H_{LK}\!(\mathbf{k} + \frac{e}{\hbar}\mathbf{A})
-
H_{LK}(\mathbf{k})\right)
}^{H_{\mathrm{orb}}(\mathbf{B})}}_{H_1(\mathbf{B})+\mathcal{O}(B^2)}
\end{equation}

where the Zeeman interaction term is defined as
\begin{equation}
    H_Z=2 \kappa \mu_B \mathbf{B} \cdot \mathbf{J} + 2 q \mu_B \left( B_x J_x^3 + B_y J_y^3 + B_z J_z^3 \right) \; 
\end{equation}
and $(\kappa,q)$ denotes the valence-band bulk Zeeman parameters~\cite{Terrazos_2021}.

We partition the Hilbert space into two subspaces: a low-energy subspace spanned by the lowest HH state $|\mathrm{HH}_0\rangle$, and a complementary excited subspace comprising all higher-energy states. The latter includes the ladder of LH states ${|\mathrm{LH}_n\rangle}$ and neglects excited $\mathrm{HH}$ subbands. Employing the usual Schrieffer–Wolff transformation, we retain the LH ladder $\{|\mathrm{LH}_n\rangle\}$ in the excited subspace and integrate them out perturbatively. In this framework, HH-LH mixing refers to the virtual coupling of the lowest $\mathrm{HH}$ subband ($|{\mathrm{HH}_0}\rangle$) to the ladder of $\mathrm{LH}$ subbands ($\{|\mathrm{LH}_n\rangle\}$) via off-diagonal Luttinger--Kohn  terms (proportional to $k_z$, $z$, and $\{z,k_z\}$).

Retaining only the linear terms in $\mathbf{B}$~\cite{Venitucci_2018, AbadilloUriel2023_PRL}, yields an effective $2\times2$ Zeeman Hamiltonian within the HH-like doublet subspace~\cite{Winkler2003_Book}
\begin{equation}
    H^{2D}_{\rm eff}=\tfrac{\mu_B}{2}\big(g_{xx}\sigma_x B_x+g_{yy}\sigma_y B_y+g_{zz}\sigma_z B_z\big) \; .
\end{equation}
To determine $g_{xx}$, we consider an in-plane magnetic field applied along the $x$-direction and evaluate the linear Zeeman splitting  
$E_{\uparrow}-E_{\downarrow}=\mu_B\, g_{xx}\, B_x$ of the HH-like doublet. The states $|\uparrow\rangle$ and $|\downarrow\rangle$ denote the two eigenstates of the resulting effective Hamiltonian. The in-plane response splits into a bulk term contribution, originating from the cubic Zeeman interaction of an unmixed $\mathrm{HH}$-like doublet, and a confinement-induced reduction term, governed by virtual coupling to $\mathrm{LH}$ subbands through the Schrieffer-Wolff transformation~\cite{Rimbach-Russ2025, Venitucci_2018}:
\begin{equation}
\begin{split}
g_{xx}(\mathbf{s})& \approx 3q- \frac{6}{m_0\Delta_{HL}}\Big[\lambda(\mathbf{s})\langle p_x^2\rangle-\lambda'(\mathbf{s})\langle p_y^2\rangle\Big]\\ 
&= 3q-{\delta g^{\text{conf}}_{xx}} \;.
\end{split}
\label{eq:gxx_s}
\end{equation}
In Eq.~\eqref{eq:gxx_s}, $\Delta_{\mathrm{HL}}$ is the vertical $\mathrm{HH}$–$\mathrm{LH}$ gap and $\langle p_{x}^{2}\rangle$ and $\langle p_{y}^{2}\rangle$ are the moments evaluated from the in-plane $\mathrm{HH}$ orbital ground-state. The coefficients $\lambda(\mathbf{s})$ and $\lambda'(\mathbf{s})$ are functionals of the vertical envelopes via the dimensionless coupling integrals $\eta_h$ and $\tilde\eta_h$: 

\begin{equation}
\lambda (\mathbf{s})  = 2\,\eta_h\,\gamma_3^2 - \kappa\,\gamma_2 + 2\,\tilde\eta_h\,\gamma_2\gamma_3 \; , 
\label{lambda_eq_def}
\end{equation}

\begin{equation}
\lambda'(\mathbf{s})  = 2\,\eta_h\,\gamma_2\gamma_3 - \kappa\,\gamma_2 + 2\,\tilde\eta_h\,\gamma_2\gamma_3 \; ,
\label{lambdap_eq_def}
\end{equation}

with 
\begin{equation}
\eta_h = 2 \Delta_{\mathrm{HL}}\,\mathrm{Im}\sum_{n=0}^{n_z^{\rm LH}-1}\frac{\langle \mathrm{HH}_0|z|\mathrm{LH}_n\rangle\langle \mathrm{LH}_n|k_z|\mathrm{HH}_0\rangle}{E^{(\mathrm{LH}_n)}-E^{(\mathrm{HH}_0)}} \; ,
\label{eq:eta}
\end{equation}
\begin{equation}
\tilde\eta_h = \Delta_{\mathrm{HL}}\mathrm{Im}\sum_{n=0}^{n_z^{\rm LH}-1}\frac{\langle \mathrm{HH}_0|\{z,k_z\}|\mathrm{LH}_n\rangle\langle \mathrm{LH}_n|\mathrm{HH}_0\rangle}{E^{(\mathrm{LH}_n)}-E^{(\mathrm{HH}_0)}} \; .
\label{eq:etat}
\end{equation}

In Eqs.~\eqref{eq:eta}--\eqref{eq:etat}, $k_z=-i\partial_z$ and $\{z,k_z\}=zk_z+k_zz$. The parameters $(\gamma_1,\gamma_2,\gamma_3)$ are the material-dependent Luttinger parameters~\cite{Terrazos_2021}. We note that in Eqs.~\eqref{eq:eta}-~\eqref{eq:etat}, we approximate the denominators by including only the vertical subband energies. This simplification is justified by the strong separation of energy scales between out-of-plane and in-plane confinement. For our parameters, $L_{x,y} \gg L_z$, making the in-plane orbital spacings $\hbar\omega_{x,y} \propto 1/L_{x,y}^2$ much smaller
than the subband splittings along $z$, $\Delta E_{(z)} \propto 1/L_z^2$. As a result, the total energy differences satisfy $E^{(\mathrm{LH}_n)}-E^{(\mathrm{HH}_0)} \simeq
E^{(\mathrm{LH}_n)}_{(z)}-E^{(\mathrm{HH}_0)}_{(z)}$, and the small in-plane contributions in the denominators are neglected.

Based on Eq.~\eqref{eq:gxx_s}, $g_{xx}$ can be modulated in two ways: (i) through the out-of-plane potential $V_\perp(z;\mathbf{s})$, which determines $\eta_h$, $\tilde\eta_h$ and $\Delta_{\mathrm{HL}}$ via the wave function envelopes $\phi^{(\mathrm{HH})}_0,\{\phi^{(\mathrm{LH})}_n\}$ and (ii) by the characteristic lengths of the in-plane wave functions, $L_x$, $L_y$, which modify $\langle p_{x,y}^2\rangle$.

\subsection{CMA-ES optimization}
\label{app:proc:opt}
We optimize the 7-segment Si concentration $\mathbf{s}=(s_1,\dots,s_7), \; s_i\in [5,20]\% \; ,$
to suppress the in-plane response, while ensuring adequate confinement of the dominant $\mathrm{HH}$ component. For each $\mathbf{s}$, the simulation pipeline (see Secs.~\ref{app:proc:vertical}–\ref{app:proc:gxx}) returns $g_{xx}(\mathbf{s})$ and the probability $P_{\rm out}(\mathbf{s})$. For clarity, we recall that the scalar cost-function is defined as: 
\begin{equation}
\label{eq:loss_proc}
\mathcal{L}(\mathbf{s}) \;=\; g_{xx}(\mathbf{s}) \;+\; 3\,\max \;\!\bigl(0\;,\;\,P_{\mathrm{out}}(\mathbf{s})-0.4\bigr) \; ,
\end{equation}
where the penalty term only provides a non-zero contribution to the cost-function when the $\mathrm{HH}$ weight outside the QW exceeds $40\%$.

We use the covariance-matrix adaptation evolution strategy (CMA-ES) in an "ask–evaluate–tell" loop~\cite{hansen2006cma} (see definition below). The algorithm maintains a multivariate Gaussian search distribution, $\mathcal{N}(\boldsymbol{\mu},\sigma^2\mathbf{C})$, over $\mathbf{s}$ and adapts its mean $\boldsymbol{\mu}$, global step size $\sigma$, and covariance $\mathbf{C}$ in each generation to iteratively guide the distribution towards regions of the search space where suitable solutions are likely to be found: 

\begin{enumerate}
\item \textbf{Ask:} Sample a population of candidates $\{\mathbf{s}_i\}$ from the current Gaussian search distribution via $\mathbf{s}_i = \boldsymbol{\mu} + \sigma\, \mathbf{C}^{1/2}\mathbf{r}_i$, where $ \mathbf{r}_i\sim\mathcal{N}(\mathbf{0},I)$,
and then project each segment to its bound.
\item \textbf{Evaluate:} For each $\mathbf{s}_i$ evaluate $\mathcal{L}(\mathbf{s}_i)$.
\item \textbf{Tell:} Rank candidates by $\mathcal{L}$ (best to worst). Update $\boldsymbol{\mu}$ by a weighted average of the top individuals; adapt $\sigma$ from the step-length statistics; update $\mathbf{C}$ so that consistently successful directions are reinforced, while ineffective ones are damped.
\end{enumerate}

The CMA-ES settings (population, initialization, termination) are listed in App.~\ref{app:num}. 

\section{Numerical implementation details}
\label{app:num}

\subsection{Strain-shifted valence-band offset}  
\label{app:num:bandedges}
The strain-shifted valence-band offset is given as follows: $U^{(\mathrm{HH})}_0=P+Q-U_0$ and $U^{(\mathrm{LH})}_0=P-Q-U_0$, with $P=-a_v(\varepsilon_{xx}+\varepsilon_{yy}+\varepsilon_{zz})$ and $Q=-(b_v/2)(\varepsilon_{xx}+\varepsilon_{yy}-2\varepsilon_{zz})$ representing the hydrostatic and shear deformation potentials, respectively. We set the Si$_{0.2}$Ge$_{0.8}$ barrier to $\SI{0}{eV}$ (outside the well). The constant term $U_0$ is the valence-band offset between Ge and Si$_{0.2}$Ge$_{0.8}$ layers. Inside the well, a piecewise Si(\%) function, containing 7 equal segments of width $L_z/N=\SI{14}{nm}/7=\SI{2}{nm}$, linearly interpolates towards the barrier value~\cite{Terrazos_2021}. This function corresponds to the potential  $W^{(j)}(z;\mathbf{s})$, introduced in Eq.~\eqref{eq:logistic}.

\subsection{Simulation domain grid, boundary conditions and out-of-plane Schrödinger equation solver}
\label{app:num:domain}
The simulation domain was defined as $z\in[z_s-\SI{40}{nm},\,z_e+\SI{60}{nm}]$ (see Fig.~\ref{fig:device_sidecut} in the main text), and discretized using a uniform grid spacing $\Delta z=\SI{0.02}{nm}$. The solutions to the out-of-plane Schr\"{o}dinger equation, Eq.~\eqref{eq:Sheq}, were constrained to obey the hard-wall boundary conditions at the domain edges: $\phi^{(j)}_n(z=-47)=\phi^{(j)}_n(z=67)=0$. We note that halving the value of $\Delta z$, or extending the lower margin by more than $\SI{40}{nm}$, does not alter the value of $g_{xx}$ by more than $10^{-5}$, demonstrating that the chosen domain and discretization parameters are well converged.

To solve the out-of-plane Schrödinger equation, Eq.~\eqref{eq:Sheq}, for each band $j \in \{\mathrm{HH},\mathrm{LH}\}$, we discretize the equations using a centered second-order finite difference approximation for the Laplacian operator. This discretization results in a tridiagonal matrix representation of the kinetic term. Consequently, solving Eq.~\eqref{eq:Sheq} reduces to computing the eigenvalues and eigenvectors of a Hermitian matrix, defined for each band as:
\begin{equation}
\begin{split}
H_j &= -\frac{\hbar^2}{2m^{(j)}_\perp (\Delta z)^2} \,\mathrm{tridiag}\;(1,-2,1) + \mathrm{diag}\;\!\big(V^{(j)}_\perp (z;\mathbf{s})\big) \; .
\end{split}
\label{discrete_Scho_z_eq}
\end{equation}

 For each subband $j\in\{\mathrm{HH},\mathrm{LH}\}$, we denote by $n^{(j)}_z$ the number of eigenpairs used in simulation. For the HH subbands, we retain only the lowest energy eigenpair ($n_z^{(\mathrm{HH})}=1$), whereas for the LH subbands, the first 50 eigenpairs are retained ($n_z^{(\mathrm{LH})}=50$). The eigenpairs are computed using the \texttt{scipy.sparse.linalg} module in Python for sparse Hermitian matrices. The numerically obtained wave functions are then normalized such that $\int |\phi^{(j)}_n(z)|^2 dz=1$.

\subsection{Evaluation of the in-plane moments}
\label{app:num:inplane}
For the $\mathrm{HH}$ ground-state orbital, we compute 
\begin{equation}
    \langle p_x^2\rangle=\hbar^2\!\int \psi^{{(\mathrm{HH})}{\ast}}_0(x)\;(-\partial_x^2)\;\psi^{(\mathrm{HH})}_0(x)\,dx \;,
\end{equation}
using a uniform grid $x\!\in[-10L_x,10L_x]$, discretized in equal steps $\Delta x=\SI{0.05}{nm}$. For $\langle p_y^2\rangle$, we proceed similarly, with $y\!\in[-10L_y,10L_y]$ and $\Delta y=\Delta x$. The integrals are evaluated using Simpson’s rule via the \texttt{simpson} function from the \texttt{scipy.integrate} module, and derivatives are computed with \texttt{numpy.gradient} with \texttt{edge\_order}=2.

\subsection{Numerical evaluation of the in-plane g-tensor components}
\label{app:num:gxx}
Given the inputs $\{\phi^{(\mathrm{HH})}_0,\phi^{(\mathrm{LH})}_n,E^{(\mathrm{HH}_0)},E^{(\mathrm{LH}_n)}\}
$, we compute $\eta_h(\mathbf{s})$, $\tilde{\eta}_h(\mathbf{s})$, which subsequently determine $\lambda (\mathbf{s})$ and $\lambda' (\mathbf{s})$ via Eq.~\eqref{lambda_eq_def} and Eq.~\eqref{lambdap_eq_def}, respectively. This procedure allows to fully determine $g_{xx}(\mathbf{s})$ through Eq.~\eqref{eq:gxx_s}. We note that the sums for $\eta_h,\tilde\eta_h$ converged for $n_z^{\mathrm{LH}}\ge 50$, as verified by checking that increasing $n_z^{\mathrm{LH}}$ from 50 to 70 alters $g_{xx}$ by less than $10^{-5}$. Based on this convergence analysis, we fixed $n_z^{\mathrm{LH}}=50$ in all our numerical simulations.

\begin{table}[t]
\caption{Physical and material parameters (Ge/Si$_{0.2}$Ge$_{0.8}$).}
\label{tab:phys-mat-params}
\begin{ruledtabular}
\begin{tabular}{lcl}
Quantity & Value & Note \\
\hline
Electron mass $m_0$ & $9.1093837\times10^{-31}\,$kg & \\
Elementary charge $e$ & $1.602176634\times10^{-19}\,$C & \\
$\hbar$ & $1.054571817\times10^{-34}\,$J\,s & \\
$U_0$ (Ge/Si$_{0.2}$Ge$_{0.8}$) & \SI{0.138}{eV} & band  offset \\
$a_v$ (Ge) & \SI{2.00}{eV} &  \\
$b_v$ (Ge) & \SI{-2.16}{eV} &  \\
Strain $(\varepsilon_{xx},\varepsilon_{yy},\varepsilon_{zz})$ & $(-0.0063,-0.0063,0.0044)$ & epitaxial  \\
$\gamma_1,\gamma_2,\gamma_3$ & $13.38,\,4.24,\,5.69$ & \\
$\kappa,\,q$ & $3.42,\,0.067$ &  \\
\end{tabular}
\end{ruledtabular}
\end{table}

\begin{table}[t]
\caption{Geometric and numerical settings}
\label{tab:num-geom}
\begin{ruledtabular}
\begin{tabular}{lcl}
Quantity & Value & Note \\
\hline
Number of segments $n_{\text{seg}}$ & 7 &   \\
QW thickness $L_z$ & \SI{14}{nm} &  \\
Interfaces $(z_s,z_e)$ & $(-\SI{7}{nm},+\SI{7}{nm})$ \\
Domain $z\in[L^{(-)}_{\mathrm{b.c.}},L^{(+)}_{\mathrm{b.c.}}]$
 & $[-\SI{47}{nm},+\SI{67}{nm}]$ &  $z$-margins \\
$z$ grid $\Delta z$ & \SI{0.02}{nm} & uniform \\
Buffer Si content $c_{\mathrm{Si}}$ & $20 \%$ \\
Vertical field $E_z$ &  \SI{0.19}{MV/m} (nominal) & Stark tilt \\
Smoothing width $\ell$ & \SI{0.2}{nm} & \ref{eq:logistic} \\
Oxide wall $L^{(+)}_{b.c.}$ & $z_e+\SI{60}{nm}$ & hard wall \\
Lower wall $L^{(-)}_{b.c.}$ & $z_s-\SI{40}{nm}$ &  hard wall \\
In-plane dot radius $L_{xy}$ & \SI{40}{nm} & isotropic \\
In-plane grid $x$ & $\Delta x=\SI{0.05}{nm}$ & $[-10L_x,10L_x]$ \\
In-plane grid $y$ & $\Delta y=\SI{0.05}{nm}$ &  $[-10L_y,10L_y]$ \\
States used & $n_z^{\mathrm{HH}}=1$, $n_z^{\mathrm{LH}}=50$ &  \\
\end{tabular}
\end{ruledtabular}
\end{table}

\subsection{Numerical convergence of simulation parameters}
\label{app:conv_parameters}

In the main text, we highlighted several parameters that can be readily adjusted for different use cases. However, applying our method to a new scenario may require preliminary numerical convergence studies to determine optimal values for certain fixed parameters in our framework. These entail the spatial grids, $\Delta z$, $\Delta x$ and $\Delta y$, the number of light-hole subbands $(n^{\mathrm{LH}}_z)$ retained in Eq.~\eqref{discrete_Scho_z_eq}, and the position of the lower Dirichlet boundary, $L^-_{b.c.}$. In our simulations, $n^{\mathrm{LH}}_z$ and $L^-_{b.c.}$ were chosen based on convergence tests for $g_{xx}$. Before running CMA-ES, in the baseline configuration, we computed $g_{xx}$ while incrementally increasing $n^{\mathrm{LH}}_z$ and deepening $L^-_{b.c.}$, allowing us to identify the smallest values beyond which further increases altered $g_{xx}$ by only $~10^{-5}$. As discussed in Sec.~\ref{app:num:domain}, a similar reasoning guided our choice for the out-of-plane grid spacing $\Delta z$.  Although these parameters remain fixed during the optimization loop, they must be carefully chosen to ensure numerical convergence of the Schrödinger equation solvers, and to correctly truncate the sums defining $\eta_h$ and $\tilde{\eta}_h$ in Eqs.~\eqref{eq:eta}-\eqref{eq:etat}.

\subsection{CMA-ES configuration and parallelization}
\label{app:num:cma}
We use \texttt{cma} Python package in an "ask–evaluate-tell" loop with population size $N_{\rm pop}=200$. The algorithm is initialized with a mean $\boldsymbol{\mu}^{(0)}=[5,10,10,10,10,10,5]\%$, a step size $\sigma^{(0)}=5\%$, an identity covariance matrix, and a projection onto the bounds $[5,20]\%$ for each segment. The evaluation step is parallelized across candidates using the \texttt{multiprocessing} Python package. In each generation, the candidate population $N_{\text{pop}}$ is partitioned across available CPU cores using the \texttt{multiprocessing} Python module, allowing parallel evaluation of the cost function. After all candidates are evaluated, the cost function values are collected and used to update the CMA-ES distribution and covariance matrix.

\section{Evolution of the HH-LH gap and in-plane g-tensor components during optimization}
\label{app:DeltaHL}
\begin{figure}[t]
  \centering
  \includegraphics[width=\columnwidth]{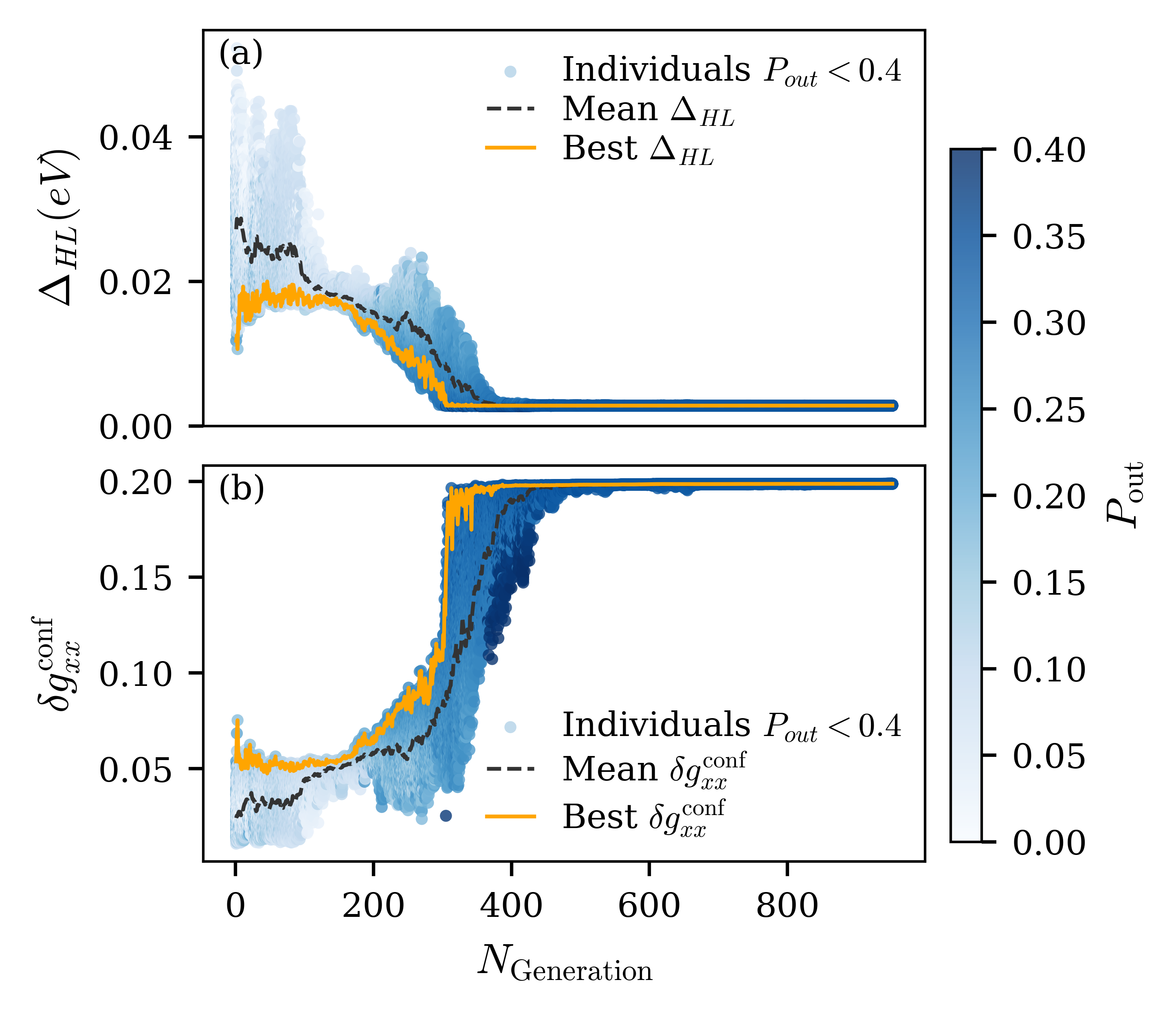}
  \caption{\textbf{Gap and confinement–induced correction through the optimization.} (a) Energy gap $\Delta_{HL}$ and (b) Confinement–induced reduction $\delta g^{\rm conf}_{xx}=3q-g_{xx}$ evaluated for each candidate across generations. The different shades of blue encode different values of $P_{\rm out}$: darker tones signal higher values of  $P_{\rm out}$. Note that we are only displaying  results satisfying $P_{\rm out}{\leq}0.4$. Solid (orange) and dashed (gray) lines show the best and mean per generation. The optimizer converges to the "double-well"-like potential, which enhances the $\mathrm{HH}$–$\mathrm{LH}$ mixing while keeping $P_{\rm out}$ bounded, reaching a minimum $\Delta_{HL}=0.00283\,\mathrm{eV}$, and minimizing $g_{xx}$.}
\label{fig:DeltaHL_gcorr_vs_generation}
\end{figure}

The suppression of $g_{xx}$ is driven by an enhanced $\mathrm{HH}$–$\mathrm{LH}$ mixing. For fixed values of $\langle p_{x,y}^2\rangle$, stronger $\mathrm{HH}$–$\mathrm{LH}$ mixing increases the contribution of $\delta g^{\rm conf}_{xx}$ in Eq.~\eqref{eq:gxx_s}, primarily by reducing $\Delta _{\mathrm{HL}}$. This trend is illustrated in Fig.~\ref{fig:DeltaHL_gcorr_vs_generation}, where the increase in $\delta g^{\rm conf}_{xx}$ across generations is accompanied by a corresponding decrease in $\Delta_{\mathrm{HL}}$.

\begin{figure}[t]
  \centering
  \includegraphics[width=\columnwidth]{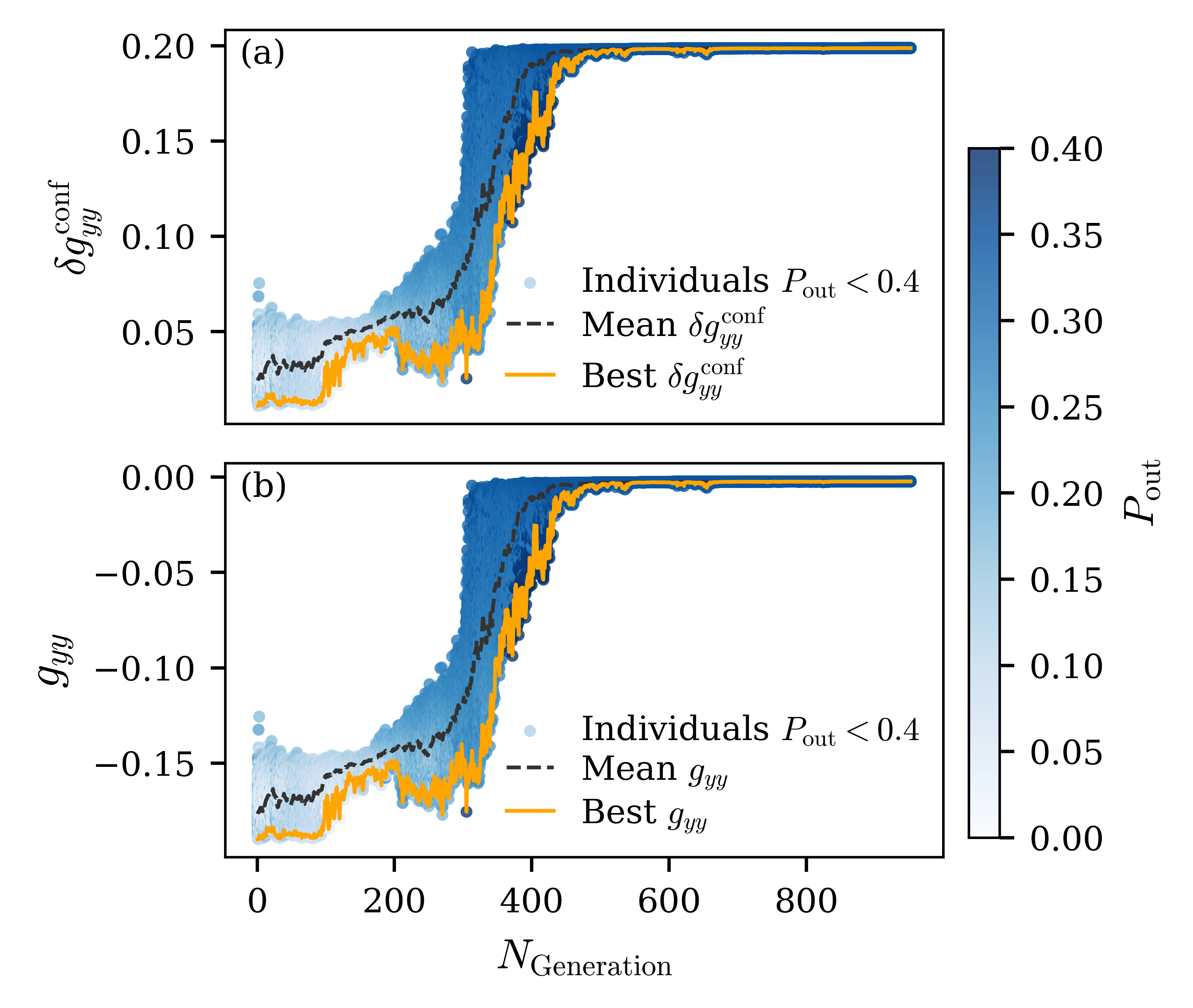}
  \caption{\textbf{Evolution of the in-plane $y$-axis Zeeman response during CMA-ES optimization.} (a) Confinement-induced reduction $\delta g^{\text{conf}}_{yy}$ and (b) the corresponding $g_{yy}$ values for all candidates with $P_{\rm out}{\leq}0.4$. The dashed curves show population means and the solid orange curves the best-of-generation. Both quantities evolve monotonically in close analogy to the corresponding $g_{xx}$ curves, demonstrating that minimizing $g_{xx}$ automatically drives $g_{yy}$ to similarly small values. This confirms that the Si-induced modification of the vertical confinement potential suppresses the entire in-plane Zeeman response}
  \label{fig:gyvsgen}
\end{figure}

In addition, the CMA-ES optimization shows an identical behavior for the in-plane component $g_{yy}$. Fig.~\ref{fig:gyvsgen} shows that both $g_{yy}$ and its confinement-induced correction, $\delta g^{\text{conf}}_{yy}$, evolve synchronously, approaching zero over the course of the optimization. This trend was to be expected, since as highlighted in the main text, we expect $g_{xx}=-g_{yy}$.

\section{Optimal in-plane g-tensor solutions across electric field strengths}
\label{app:gvsE}

\begin{figure}[t]
  \centering
  \includegraphics[width=\columnwidth]{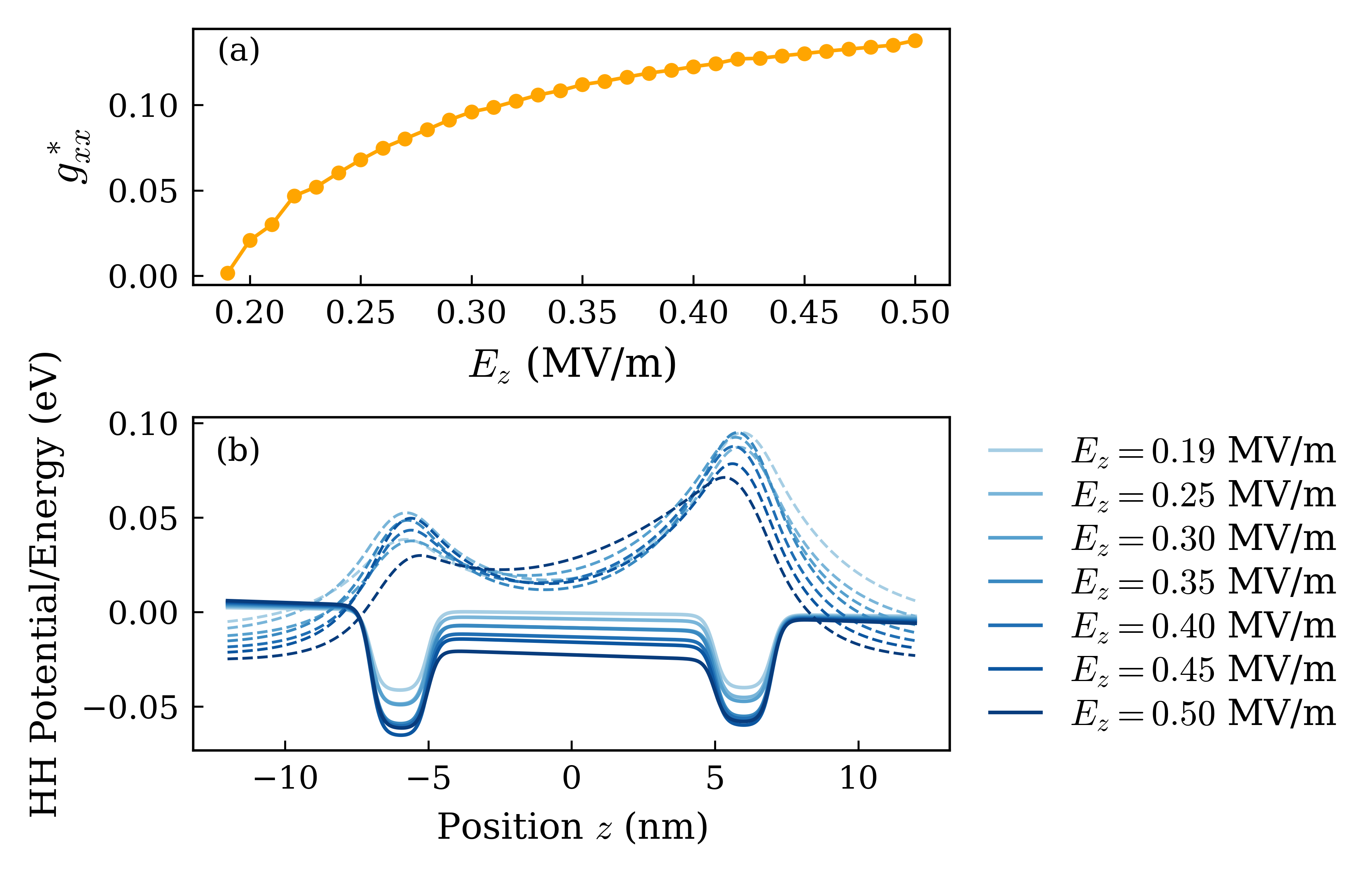}
  \caption{\textbf{Dependence of optimized $g^*_{xx}$ on the vertical electric field.}
  (a) Best-achieved values of $g^*_{xx}$ (orange markers), obtained from independent CMA-ES optimizations at each $E_z$. A monotonic increase is observed as the Stark tilt reduces confinement-induced suppression. (b) Optimized HH potentials $V_{\mathrm{HH}_0}(z;\mathrm{s}_{\text{opt}})$ (solid curves) and
  ground-state densities $|\phi_{\mathrm{HH}0}(z)|^2$ (dashed curves) for fields spanning $\SIrange{0.19}{0.50}{MV/m}$. With increasing $E_z$ the optimizer deepens the confining wells, lowers the intermediate plateau, and consistently
  creates a slightly deeper left well than right well. These adjustments compensate for the stronger Stark tilt and prevent the $\mathrm{HH}$ envelope from drifting into the oxide. The resulting reduction in $\mathrm{HH}$--$\mathrm{LH}$
  overlap weakens the virtual mixing that suppresses $g_{xx}$, producing the trend in panel~(a). Despite these quantitative adaptations, the qualitative
  double-well Si profile remains unchanged across all fields, demonstrating the intrinsic robustness of the geometric suppression mechanism.}
\label{fig:gvsE}
\end{figure}

To examine how the optimal $g_{xx}$ solution depends on the strength of the applied vertical electric field, $E_z$, we repeated the optimization loop across different electric field strengths. The resulting trend is shown in Fig.~\ref{fig:gvsE}. Panel~(a) shows the best-achieved $g_{xx}$ values, labeled $g^*_{xx}$, for different $E_z$ values. Panel~(b) provides structural insight by showing the optimized HH potentials $V^{\mathrm{HH}_0}_\perp(z;\mathrm{s}_{\text{opt}})$ (solid curves), together with the corresponding HH ground-state densities (dashed curves), $|\phi^{\mathrm{HH}}_0(z)|^2$, for fields ranging from $\SI{0.19}{MV/m}$ to $\SI{0.50}{MV/m}$. As the electric field increases, the Stark tilt pushes the $\mathrm{HH}$ envelope towards the oxide barrier. In response, the optimizer reshapes the silicon composition profile so that: (i) the depths of the confining wells increase with $E_z$; (ii) the inter-well plateau is lowered; and (iii) the left-hand well is consistently slightly deeper than the right-hand one. These adjustments counteract the increasing Stark tilt and ensure that the $\mathrm{HH}$ ground state remains well confined inside the quantum well. Hence, across all electric field values, CMA-ES converges to the same qualitative double-well Si pattern; only the depths of the wells and the plateau are modulated to offset the changing Stark tilt. Notably, $g^*_{xx}$ gradually increases with increasing applied electric field.

\section{The behavior of the optimized in-plane g-tensor components for different lateral confinement lengths}
\label{app:gvsLxy}

\begin{figure}[t]
  \centering
  \includegraphics[width=\columnwidth]{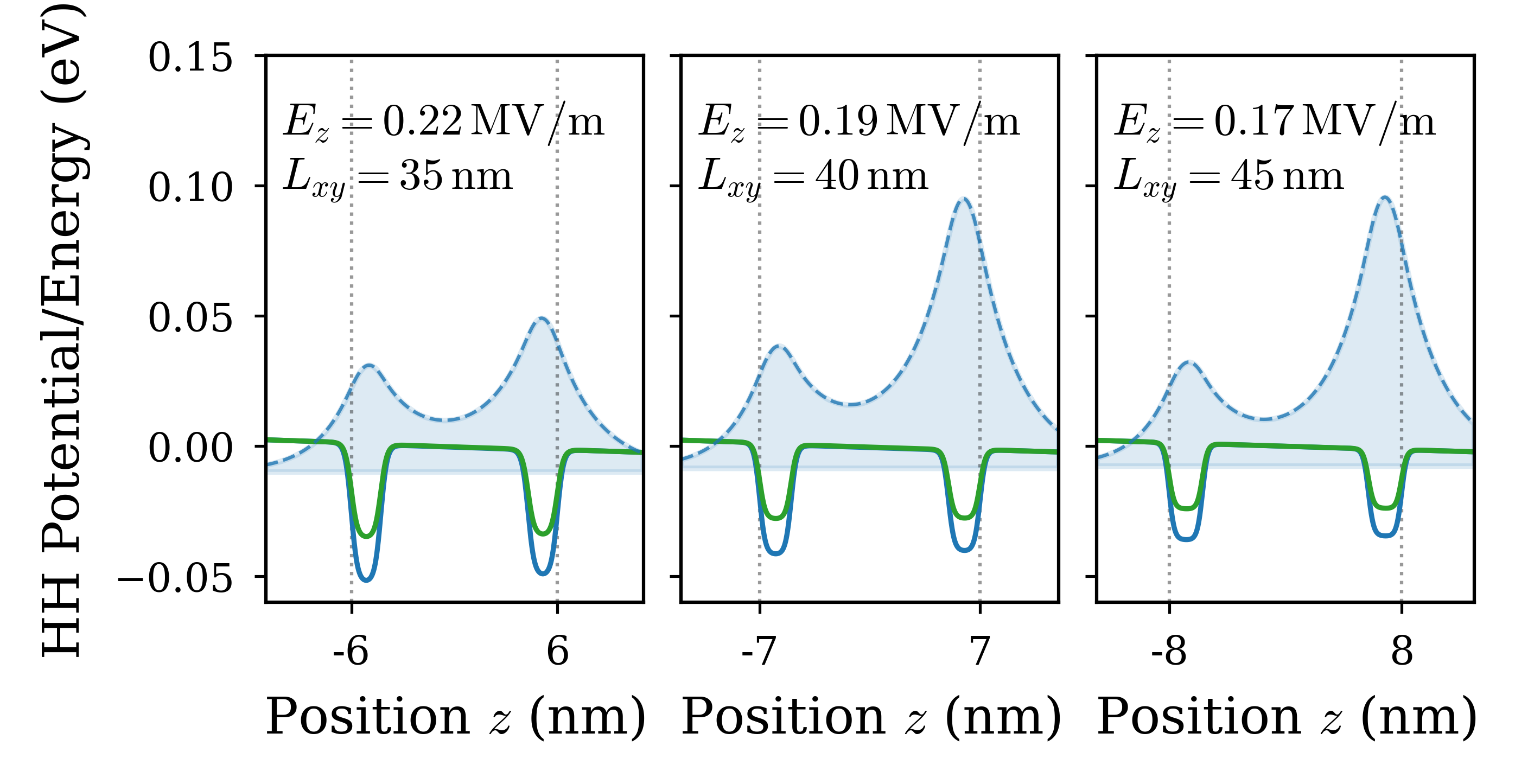}
  \caption{\textbf{Effect of lateral dot size on the optimized vertical potential.} Vertical HH (blue) and LH (green) potentials and ground-state HH envelopes for three lateral confinement lengths $L_{x,y} = \{35,40,45\}\,\mathrm{nm}$
 (left to right), each obtained from an independent CMA-ES optimization. All three structures achieve a near-zero in-plane Zeeman response $g_{xx} \sim 10^{-3}$, but at different vertical fields; larger dots require smaller $E_z$ to reach the gapless single-spin qubit  regime, because the in-plane kinetic contribution $\delta g^{\text{conf}}_{xx} \propto 1/L^2_{x,y}$ decreases with increasing lateral size. The optimized Si profiles preserve the same double-well pattern across all three cases, while the well depths adapt to counteract the Stark tilt and keep the HH envelope inside the quantum well.}
\label{fig:Lxyvsg}
\end{figure}

The lateral confinement lengths, $L_{x}$ and $L_y$, control the in-plane kinetic energy moments, $\langle p_{x}^2\rangle$ and $\langle p_y^2 \rangle$, respectively, which in turn set the strength of the confinement-induced correction $\delta g^{\text{conf}}_{xx}$ in Eq.~\eqref{eq:gxx_formal}. Since $\langle p_{x,y}^2\rangle \propto 1/L^2_{x,y}$, smaller dots naturally produce a larger kinetic contribution and therefore a stronger suppression of the in-plane Zeeman response. Fig.~\ref{fig:Lxyvsg} shows how this mechanism interacts with the vertical electric field. Each panel shows an independently optimized Si profile for a different lateral size $L_{x,y} = \{35,40,45\}\,\mathrm{nm}$, together with the applied vertical potential that guarantees attaining $g_{xx}\sim 10^{-3}$. Although all three optimizations achieve a similarly small $g_{xx}$ value, they do that at different values of the electric field. Larger dots require a smaller $E_z$ to reach the same $g_{xx}$ value, reflecting the fact that the lateral kinetic contribution weakens with increasing $L_{x,y}$. In other words, as the dot becomes wider, $\delta g^{\text{conf}}_{xx}$ diminishes and the system must rely more heavily on vertical HH-LH mixing controlled by Stark tilt, to enter the regime where $g_{xx} \approx 0$. Accordingly, the optimal field shifts downward with increasing $L_{x,y}$, exactly as seen in Fig.~\ref{fig:Lxyvsg}.

Interestingly, the optimized Si concentration retains the same double-well pattern across all three cases. What changes is the depth of the engineered double-well profile: for smaller $L_{x,y}$, the optimizer adjusts the depth of the left and right wells to keep the HH envelope confined against the Stark tilt. The left well remains slightly deeper in all cases, compensating for the field that tends to push the HH towards the oxide. This consistent pattern shows that the suppression mechanism \textemdash achieved by modifying the vertical confinement potential through the added Si concentration profile \textemdash is robust, rather than a finely tuned solution that depends on specific parameter choices.

\begin{figure}[t]
  \centering
  \includegraphics[width=0.8\columnwidth]{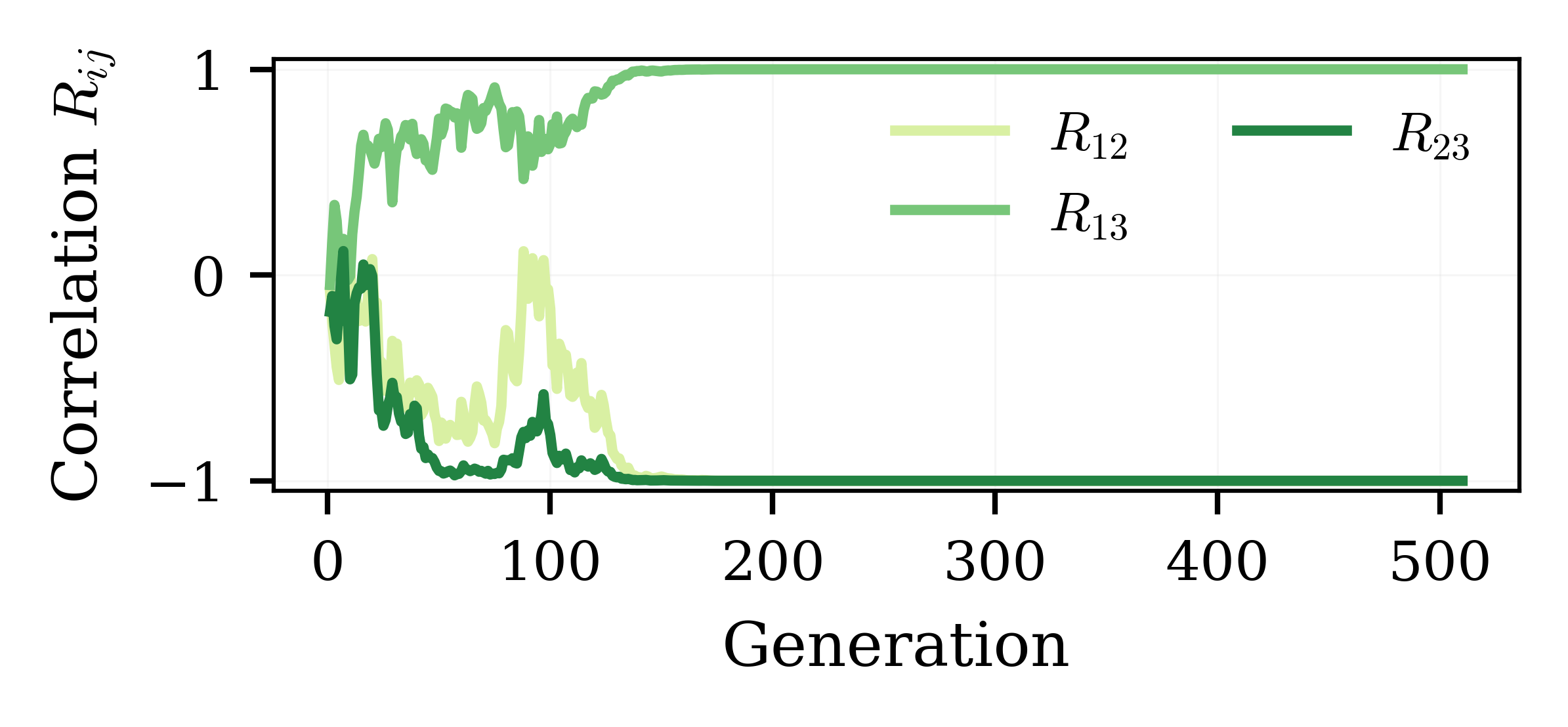}
  \caption{\textbf{Correlation evolution of the reduced three-parameter Si profile.}
  Evolution of the pairwise correlation coefficients $R_{ij}$ between ($v_1$, $v_2$, $v_3$) representing the left edge, central plateau, and right edge of the Si composition profile, respectively, as the CMA-ES optimization progresses. The strong positive $R_{13}$ between the two edges, combined with the negative correlations $R_{12}$ and $R_{23}$ between the edges and the plateau, indicates a coordinated trade-off: the two edges evolve symmetrically, while the central region compensates in the opposite direction to sustain optimal $\mathrm{HH}$–$\mathrm{LH}$ mixing and minimize $g_{xx}$.}
  \label{fig:corr}
\end{figure}

\section{Correlations between Si-profile controls}
\label{app:corr}
To gain a better understanding of the resulting optimal shape for the out-of-plane confinement potential, we carried out a reduced-parameter CMA-ES run in which the Si composition inside the Ge well is represented by only three parameters:
$v_1$ (left edge), $v_2$ (central plateau), and $v_3$ (right edge). This compact parametrization retains the essential structural degrees of freedom that shape the out-of-plane potential, making it possible to visualize how the optimization coordinates adjustments across the well to minimize $g_{xx}$.

At each generation, we transformed the evolving CMA covariance matrix, $C$, into the normalized correlation matrix $R_{ij}=C_{ij} /(\sigma_i\sigma_j)$~\cite{hansen2006cma}, where $\sigma_i$ are the marginal standard deviations of each variable. We recall that candidates that lead to poor confinement ($P_{\mathrm{out}}>0.4$) are penalized in Eq.~\eqref{eq:loss_proc} and therefore have negligible influence on the correlation trends.

The correlation trajectories in Fig.~\ref{fig:corr} exhibit a consistent pattern: the two interface parameters ($v_1$ and $v_3$) are strongly positively correlated, whereas both show negative correlation with the central plateau $v_2$. This means that the optimizer adjusts the two edges cooperatively—when one interface deepens or shallows, the other tends to move in the same direction—while the central plateau compensates in the opposite manner. This dynamic reflects a physically intuitive balancing act: the optimal solution preserves an approximately symmetric double-well-like structure to regulate the overall confinement and the $\mathrm{HH}$–$\mathrm{LH}$ mixing. The result is an optimized potential that maintains the double-well pattern and maximizes the confinement-induced suppression of $g_{xx}$.

\section{Optimized Si profile for different well thickness sizes}
\label{app:gvsLz}

\begin{figure}[t]
  \centering
  \includegraphics[width=\columnwidth]{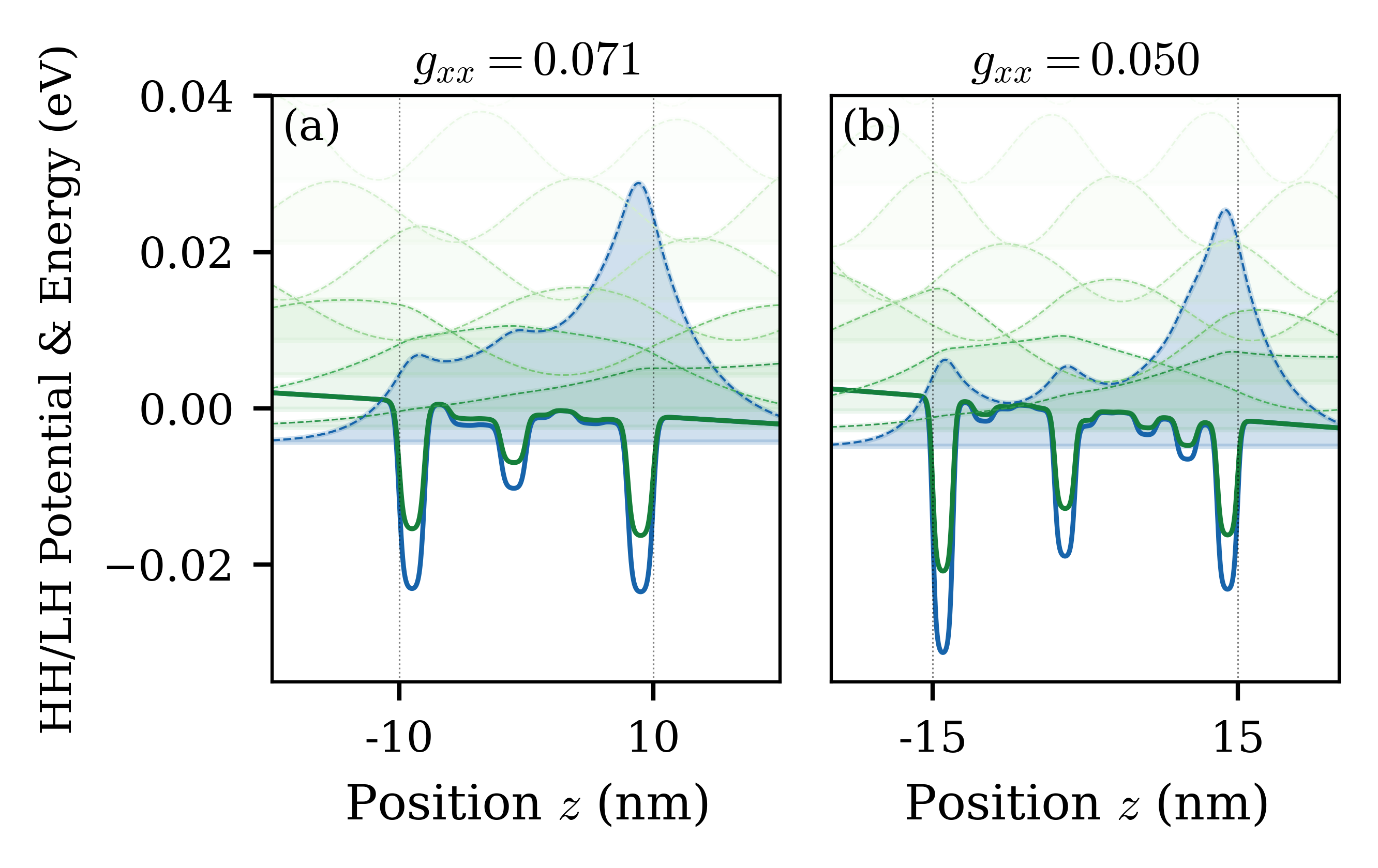}
  \caption{\textbf{Optimized vertical confinement for thicker wells.}
Optimized out-of-plane confinement profiles for (a) $L_z=\SI{20}{nm}$ (10 segments) and (b) $L_z=\SI{30}{nm}$ (15 segments). Solid lines show the HH and LH potentials, while dashed curves illustrate the corresponding probability densities. Vertical dashed lines indicate the nominal well boundaries at $z=\pm L_z/2$. In both cases, the optimizer produces two pronounced Si-induced wells near the left/right sides, yielding a bimodal wave function with a shallow central well that is consistent with an optimization-driven increase of left--right overlap (tunneling bridge) when the two lobes would otherwise become too weakly coupled.} \label{fig:Lz_profiles}
\end{figure}

When increasing the well thickness from the nominal $L_z=\SI{14}{nm}$ to $L_z=\SI{20}{nm}$ and $\SI{30}{nm}$ (discretized into 2-nm Si-profile segments), the optimizer consistently retains two deep Si-induced wells near the left/right sides (see Fig.~\ref{fig:Lz_profiles}), which favor a bimodal probability density with two main lobes (i.e., two density peaks). However, for larger $L_z$ an additional shallow central well emerges. A natural interpretation is that this middle feature acts as a tunneling bridge: as the separation between the two dominant lobes increases, their overlap is suppressed by the exponential decay of the wave-function amplitude across the barrier separating them. In a barrier plateau, the wave-function amplitude decays exponentially with a characteristic decay length, causing the overlap to decrease rapidly with increasing separation. Therefore, when $L_z$ becomes large enough that the lobe separation exceeds the relevant decay length, the optimizer can recover finite coupling by inserting a shallow intermediate well that restores overlap between left/right lobes. This increased overlap modifies the spatial extent of the relevant wave-function envelopes and thereby affects the overlap-dependent renormalization of the $g_{xx}$.

\FloatBarrier

\section{Out-of-plane g-tensor component under the optimized Si pattern}
\label{app:gzz}

\begin{figure}[t]
  \centering
  \includegraphics[width=\columnwidth]{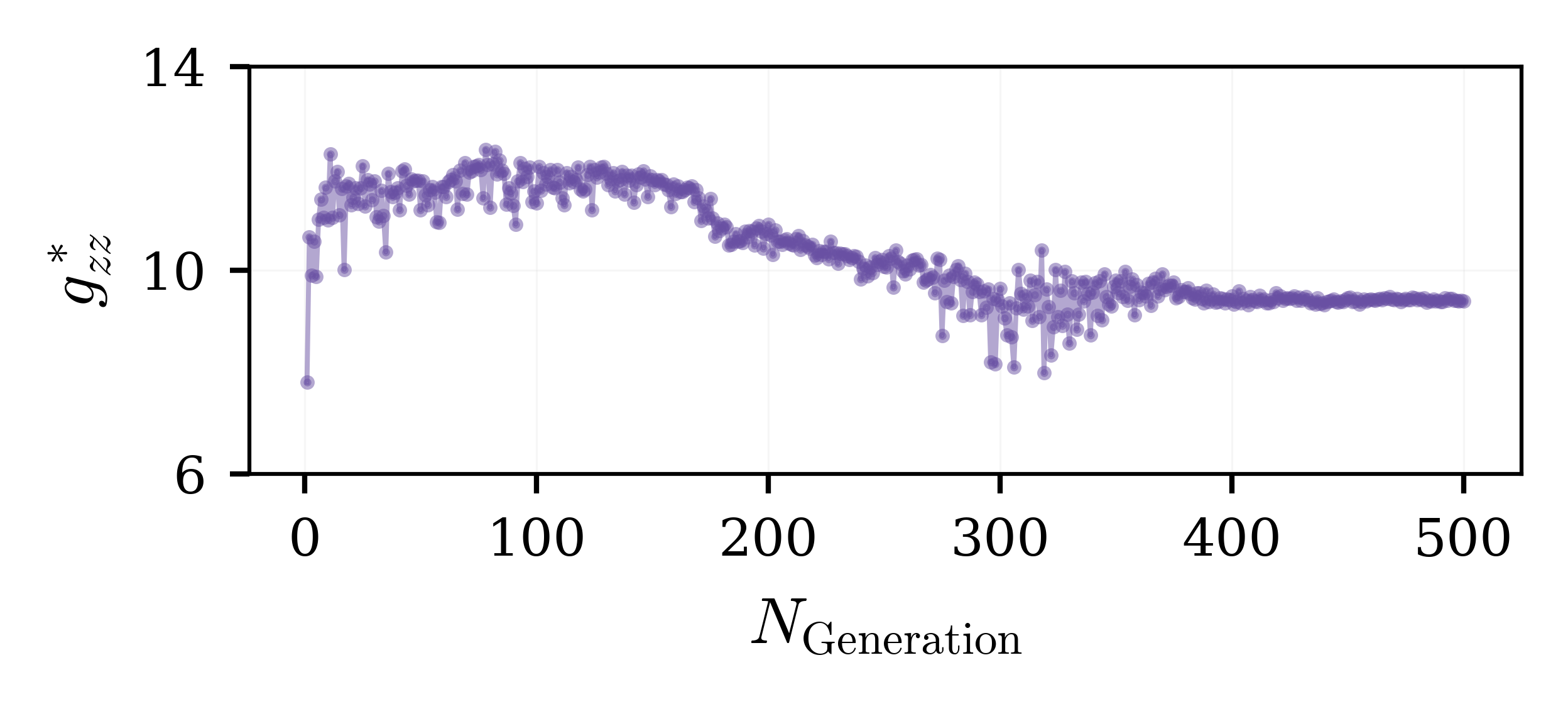}
  \caption{\textbf{Evolution of the out-of-plane Zeeman response during optimization.} Computed values of the $\mathrm{HH}$ $g_{zz}$ component related to $g^*_{xx}$ across CMA-ES generations for the optimized seven-segment Si profile ($L_z=\SI{14}{nm}$, $L_{x,y}=\SI{40}{nm}$, $E_z=\SI{0.19}{MV/m}$). Although the optimization targets $g_{xx}$, the resulting increase in $\mathrm{HH}$–$\mathrm{LH}$ mixing also produces a moderate reduction of $g_{zz}$ from $\sim16$ to $\sim9$. This correlated behavior reflects that the same virtual coupling mechanism governing in-plane suppression also renormalizes the longitudinal Zeeman response, yielding a more isotropic $g$-tensor in the optimized well.}
\label{fig:gzz}
\end{figure}

The same $\mathrm{HH}$–$\mathrm{LH}$ mixing that drives the suppression of the HH-like doublet in-plane response, $g_{xx}$, also modifies the out-of-plane component $g_{zz}$. Within the Luttinger–Kohn framework, the HH-like doublet $g_{zz}$ factor can be expressed as $g_{zz} = g^{\text{bulk}}_{zz}- \delta g^{\text{conf}}_{zz}$ \cite{Rimbach-Russ2025}: 
\begin{equation}
\label{eq:gzz_def}
g_{zz} \; = \; 6\,\kappa \;+\; \frac{27}{2}\,q \; -\; 2\,\gamma_h \;=\; g^{\text{bulk}}_{zz} - \delta g^{\text{conf}}_{zz},
\end{equation}
where the first two terms give rise to its bulk value, and the confinement correction term $\gamma_h$ arises from virtual coupling to $\mathrm{LH}$ subbands and is captured by $\gamma_h$~\cite{Rimbach-Russ2025, Venitucci_2018}. This latter term is defined as

\begin{equation}
\gamma_h \;=\; \frac{6\,\gamma_3^{\,2}\hbar^{2}}{m_0}
\sum_{n=0}^{n_z^{\rm LH}-1}\frac{\big|\langle \mathrm{HH}_0 \lvert k_z \rvert \mathrm{LH}_n\rangle\big|^{2}}
{E^{(\mathrm{LH}_n)}-E^{(\mathrm{HH}_0)}} \; .
\label{eq:gammah_def}
\end{equation}

A smaller subband gap $E_{\mathrm{LH},n}-E_{\mathrm{HH},0}$, or larger overlap $\langle \mathrm{HH}_0|k_z|\mathrm{LH}_n\rangle$, enhances $\gamma_h$, and since it enters Eq.~\eqref{eq:gzz_def} with a negative sign, stronger mixing reduces $g_{zz}$. Numerically, the optimized double-well-like Si profile, while primarily minimizing $g_{xx}$, also lowers $g_{zz}$ from about $\sim16$ in the unpatterned well to $\sim9$ (Fig.~\ref{fig:gzz}). This reduction reflects the same underlying mechanism: increased $\mathrm{HH}$–$\mathrm{LH}$ admixture weakens the pure $J_z=\pm3/2$ character of the HH-like doublet, yielding a softer and more isotropic Zeeman response without losing HH confinement.

\FloatBarrier
\bibliographystyle{apsrev4-2}
\bibliography{references}

\end{document}